\RequirePackage{fix-cm}
\documentclass[twocolumn,epjc3]{svjour3} 
\smartqed % flush right qed marks, e.g. at end of proof
\RequirePackage{graphicx}
\RequirePackage{verbatim}
\usepackage{lineno}
\usepackage{hyperref}
\usepackage{cite}
\begin{document}
%\linenumbers
%\begin{document}
\title{New limits on the Pauli forbidden transitions in $\mathrm{^{12}C}$ nuclei obtained with the complete Borexino dataset}
 \author{
 D.~Basilico\thanksref{Milano}
 \and
 G.~Bellini\thanksref{Milano}
 \and
 J.~Benziger\thanksref{PrincetonChemEng}
 \and
 R.~Biondi\thanksref{LNGS,GSSI} % \author[LNGS]{R.~Biondi\fnref{GSSI}}
 \and
 B.~Caccianiga\thanksref{Milano}
 \and
% F.~Calaprice\thanksref{Princeton}
 %\and
 A.~Caminata\thanksref{Genova}
 \and
 A.~Chepurnov\thanksref{Lomonosov}
 \and
 D.~D'Angelo\thanksref{Milano}
 \and
 A.~Derbin\thanksref{Peters,Kurchatov}
 \and
 A.~Di Giacinto\thanksref{LNGS}
 \and
 V.~Di Marcello\thanksref{LNGS}
 \and
 X.F.~Ding\thanksref{Princeton,IHEP} %\author[Princeton]{X.F.~Ding\fnref{IHEP}}
 \and
 A.~Di Ludovico\thanksref{Princeton,LNGSG} 
 \and
 L.~Di Noto\thanksref{Genova}
 \and
 I.~Drachnev\thanksref{Peters}
 \and
 D.~Franco\thanksref{APC}
 \and
 C.~Galbiati\thanksref{Princeton}
 \and
 C.~Ghiano\thanksref{LNGS}
 \and
 M.~Giammarchi\thanksref{Milano}
 \and
 A.~Goretti\thanksref{LNGSG,Princeton} 
 \and
 M.~Gromov\thanksref{Lomonosov,Dubna}
 \and
 D.~Guffanti\thanksref{Mainz,Bicocca}
 \and
 Aldo~Ianni\thanksref{LNGS}
 \and
 Andrea~Ianni\thanksref{Princeton}
 \and
 A.~Jany\thanksref{Krakow}
 \and
 V.~Kobychev\thanksref{Kiev}
 \and
 G.~Korga\thanksref{London,Atomki}
 \and
 S.~Kumaran\thanksref{Juelich,RWTH,CALI}
 \and
 M.~Laubenstein\thanksref{LNGS}
 \and
 E.~Litvinovich\thanksref{Kurchatov,Kurchatovb}
 \and
 P.~Lombardi\thanksref{Milano}
 \and
 I.~Lomskaya\thanksref{Peters}
 \and
 L.~Ludhova\thanksref{Juelich,RWTH,GSIMAINZ} %\author[Juelich,RWTH]{L.~Ludhova\fnref{GSIMAINZ}}
 \and
 I.~Machulin\thanksref{Kurchatov,Kurchatovb}
 \and
 J.~Martyn\thanksref{Mainz}
 \and
 E.~Meroni\thanksref{Milano}
 \and
 L.~Miramonti\thanksref{Milano}
 \and
 M.~Misiaszek\thanksref{Krakow}
 \and
 V.~Muratova\thanksref{Peters}
 \and
% R.~Nugmanov\thanksref{Kurchatov,Kurchatovb}
 %\and
 L.~Oberauer\thanksref{Munchen}
 \and
 V.~Orekhov\thanksref{Mainz}
 \and
 F.~Ortica\thanksref{Perugia}
 \and
 M.~Pallavicini\thanksref{Genova}
 \and
 L.~Pelicci\thanksref{Juelich,RWTH,MI}
 \and
 \"O.~Penek\thanksref{Juelich,BOS} %\author[Juelich]{\"O.~Penek\fnref{BOS}}
 \and
 L.~Pietrofaccia\thanksref{Princeton,LNGSG}
 \and
 N.~Pilipenko\thanksref{Peters}
 \and
 A.~Pocar\thanksref{UMass}
 \and
 G.~Raikov\thanksref{Kurchatov}
 \and
 M.T.~Ranalli\thanksref{LNGS}
 \and
 G.~Ranucci\thanksref{Milano}
 \and
 A.~Razeto\thanksref{LNGS}
 \and
 A.~Re\thanksref{Milano}
 \and
 N.~Rossi\thanksref{LNGS}
 \and
 S.~Sch\"onert\thanksref{Munchen}
 \and
 D.~Semenov\thanksref{Peters}
 \and
 G.~Settanta\thanksref{Juelich,ISPRA}
 \and
 M.~Skorokhvatov\thanksref{Kurchatov,Kurchatovb}
 \and
 A.~Singhal\thanksref{Juelich,RWTH,WSS}
 \and
 O.~Smirnov\thanksref{Dubna}
 \and
 A.~Sotnikov\thanksref{Dubna}
 \and
 R.~Tartaglia\thanksref{LNGS}
 \and
 G.~Testera\thanksref{Genova}
 \and
 E.~Unzhakov\thanksref{Peters}
 \and
 A.~Vishneva\thanksref{Dubna}
 \and
 R.B.~Vogelaar\thanksref{Virginia}
 \and
 F.~von~Feilitzsch\thanksref{Munchen}
 \and
 M.~Wojcik\thanksref{Krakow}
 \and
 M.~Wurm\thanksref{Mainz}
 \and
 S.~Zavatarelli\thanksref{Genova}
 \and
 K.~Zuber\thanksref{Dresda}
 \and
 G.~Zuzel\thanksref{Krakow}
 \\ (The Borexino collaboration){}
 }
 
\thankstext{IHEP}{Present address: IHEP Institute of High Energy Physics, 100049 Beijing, China}
\thankstext{LNGSG}{Present address: INFN Laboratori Nazionali del Gran Sasso, 67010 Assergi (AQ), Italy}
\thankstext{Bicocca}{Present address: Dipartimento di Fisica, Universit\`a degli Studi e INFN Milano-Bicocca, 20126 Milano, Italy}
\thankstext{CALI}{Present address: Department of Physics and Astronomy, University of California, Irvine, California, USA}
\thankstext{GSIMAINZ}{Present address: GSI Helmholtzzentrum f\"ur Schwerionenforschung GmbH, Planckstr. 1, 64291 Darmstadt, Germany and Institute of Physics and EC PRISMA+, Johannes Gutenberg Universit\"at Mainz, Mainz, Germany}
\thankstext{MI}{Dipartimento di Fisica, Universit\`a degli Studi e INFN, 20133 Milano, Italy}
\thankstext{ISPRA}{Present address: Istituto Superiore per la Protezione e la Ricerca Ambientale, 00144 Roma, Italy}
\thankstext{WSS}{Present address: WSS-Forschungszentrum catalaix, RWTH Aachen University, Worringerweg 2, 52074 Aachen}
\thankstext{BOS}{Present address: Boston University, College of Arts and Sciences, Department of Physics, 02215 Boston, MA, USA}

% \thankstext{MPIH}{Present address: Max-Planck-Institut f\"ur Kernphysik, 69117 Heidelberg, Germany}
%\thankstext{IHEP}{Present address: IHEP Institute of High Energy Physics, 100049 Beijing, China}
%\thankstext{LNGSG}{Present address: INFN Laboratori Nazionali del Gran Sasso, 67010 Assergi (AQ), Italy}
%\thankstext{Bicocca}{Present address: Dipartimento di Fisica, Universita degli Studi e INFN Milano-Bicocca, 20126 Milano, Italy}
%\thankstext{CALI}{Present address: Department of Physics and Astronomy, University of California, Irvine, California, USA}
%\thankstext{GSI}{Present address: GSI Helmholtzzentrum f\"r Schwerionenforschung GmbH, 64291 Darmstadt, Germany}
%\thankstext{ISPRA}{Present address: Istituto Superiore per la Protezione e la Ricerca Ambientale, 00144 Roma, Italy}

% \thankstext{Berkeley}{Present address: University of California, Berkeley, Department of Physics, CA 94720, Berkeley, USA}
 % \thankstext{Milanoa}{Present address: Dipartimento di Fisica, Universit\'a degli Studi e INFN Milano-Bicocca, 20126 Milano, Italy}
 %\thankstext{Padova}{Present address: Dipartimento di Fisica e Astronomia dell' Universit\'a di Padova and INFN Sezione di Padova, Padova, Italy}
 %\thankstext{Romaa}{Present address: Istituto Superiore per la Protezione e la Ricerca Ambientale, 00144 Roma, Italy}

% \institute{\bf{The Borexino Collaboration}}
% \and
 \institute{Dipartimento di Fisica, Universit\`a degli Studi e INFN, 20133 Milano, Italy\label{Milano}
 \and
 Chemical Engineering Department, Princeton University, Princeton, NJ 08544, USA\label{PrincetonChemEng}
 \and
 INFN Laboratori Nazionali del Gran Sasso, 67010 Assergi (AQ), Italy\label{LNGS}
 \and
 Physics Department, Princeton University, Princeton, NJ 08544, USA\label{Princeton}
 \and
 Dipartimento di Fisica, Universit\`a degli Studi e INFN, 16146 Genova, Italy\label{Genova}
 \and
 Lomonosov Moscow State University Skobeltsyn Institute of Nuclear Physics, 119234 Moscow, Russia\label{Lomonosov}
 \and
 St. Petersburg Nuclear Physics Institute NRC Kurchatov Institute, 188350 Gatchina, Russia\label{Peters}
 \and
 National Research Centre Kurchatov Institute, 123182 Moscow, Russia\label{Kurchatov}
 \and
 APC, Universit\'e de Paris, CNRS, Astroparticule et Cosmologie, Paris F-75013, France\label{APC}
 \and
 Gran Sasso Science Institute, 67100 L'Aquila, Italy\label{GSSI}
 \and
 Joint Institute for Nuclear Research, 141980 Dubna, Russia\label{Dubna}
 \and
 Institute of Physics and Excellence Cluster PRISMA+, Johannes Gutenberg-Universit\"at Mainz, 55099 Mainz, Germany\label{Mainz}
 \and
 M.~Smoluchowski Institute of Physics, Jagiellonian University, 30348 Krakow, Poland\label{Krakow}
 \and
 Institute for Nuclear Research of NAS Ukraine, 03028 Kyiv, Ukraine\label{Kiev}
 \and
 Department of Physics, Royal Holloway, University of London, Egham, Surrey, TW20 OEX, UK\label{London}
 \and
 Institute of Nuclear Research (Atomki), Debrecen, Hungary\label{Atomki}
 \and
 Institut f\"ur Kernphysik, Forschungszentrum J\"ulich, 52425 J\"ulich, Germany\label{Juelich}
 \and
 RWTH Aachen University, 52062 Aachen, Germany\label{RWTH}
 \and
 National Research Nuclear University MEPhI (Moscow Engineering Physics Institute), 115409 Moscow, Russia\label{Kurchatovb}
 \and
 Physik-Department, Technische Universit\"at M\"unchen, 85748 Garching, Germany\label{Munchen}
 \and
 Dipartimento di Chimica, Biologia e Biotecnologie, Universit\`a degli Studi e INFN, 06123 Perugia, Italy\label{Perugia}
 \and
 Amherst Center for Fundamental Interactions and Physics Department, UMass, Amherst, MA 01003, USA\label{UMass}
 \and
 Physics Department, Virginia Polytechnic Institute and State University, Blacksburg, VA 24061, USA\label{Virginia}
 \and
 Department of Physics, Technische Universit\"at Dresden, 01062 Dresden, Germany\label{Dresda}
 }
%\collaboration{\bf{The Borexino collaboration}}
\date{Received: date / Accepted: date}
\maketitle
\begin{abstract}
The Pauli exclusion principle (PEP) was tested for nucleons in $\mathrm{^{12}C}$ nuclei using the Borexino dataset from 2007 to 2021. %the complete Borexino detector data.
The approach consists of searching for $\gamma$-quanta, neutrons, protons, as well as electrons and positrons emitted in non-Paulian transitions of nucleons from the $1P_{3/2}$ shell to the filled $1S_{1/2}$ shell. 
Due to the uniquely low background level, the large mass, and long measurement time of the Borexino detector, the most stringent experimental constraints to date on the lifetime of the $\mathrm{^{12}C}$ nucleus with respect to PEP-forbidden transitions were obtained:
%The Pauli exclusion principle (PEP) has been tested for nucleons ($n,p$) in $^{12}C$ with the Borexino detector.
%The approach consists of a search for $\gamma$, $n$, $p$ and $\beta^\pm$ emitted in a non-Paulian transition of 1$P_{3/2}$- shell nucleons to the filled 1$S_{1/2}$ shell in nuclei. 
%Due to the extremely low background and the large mass (278 t) of the Borexino detector, the following most stringent up-to-date experimental bounds on PEP violating transitions of nucleons have been established:
$\tau({^{12}\mathrm{C}}\rightarrow{^{12}\widetilde{\mathrm{C}}}+\gamma) \geq {1.1\times10^{32}}$ y, $\tau({^{12}\mathrm{C}}\rightarrow{^{11}\widetilde{\mathrm{B}}}+ p) \geq {1.0\times10^{31}}$ y, $\tau({^{12}\mathrm{C}}\rightarrow{^{11}\widetilde{\mathrm{C}}}+ n) \geq 2.0 \times 10^{31}$ y, $\tau({^{12}\mathrm{C}}\rightarrow{^{12}\widetilde{\mathrm{N}}}+ e^- + \overline{\nu}_e) \geq 6.4 \times 10^{30}$ y and $\tau({^{12}\mathrm{C}}\rightarrow{^{12}\widetilde{\mathrm{B}}}+ e^+ + \nu_e) \geq 6.6 \times 10^{30}$ y (90\% C.L.). 
The upper limits on the relative strengths for the non-Paulian electromagnetic, strong, and weak transitions have been obtained: $\delta^2_{\gamma}\leq 1.0\times 10^{-57}$, $\delta^2_{N}\leq 7.0\times 10^{-61}$ and $\delta^2_{\beta}\leq 9.6\times 10^{-36}$, all at 90\% C.L..
\end{abstract}
%\pacs{11.30.-j, 24.80.+y, 23.20.-g, 27.20.+n} 
%\keywords {Pauli exclusion principle; Borexino; rare processes}
\maketitle

\section{Introduction}
%%%%%%%%%%%%%%%%%%%%%%%%%%%%%%%%%%%%%%%%%%%%%%%%%%%%%%%%%%%%%%%%%%%%%%%%%%%%%%%%%%%%%%%%%%%%%%%
The Pauli exclusion principle (PEP) is one of the cornerstones of physics and chemistry. %, although the understanding of the reasons for the existence of the PP remains unsatisfactory.
The principle was formulated by W.~Pauli in 1925 before the creation of quantum mechanics and in its original form postulated that in the case of Bohr atom "there can never be two or more equivalent electrons in an atom"~\cite{Pau25}. 
Within the framework of quantum mechanics, it was established that for two identical electrons, the total wave function is antisymmetric with respect to the permutation of electrons.
Relativistic quantum field theory with the anticommutativity of the fermion creation and annihilation operators inevitably leads to the Pauli exclusion principle for systems of identical fermions.
The spin-statistics theorem relates the spin of a particle and the statistics that governs an ensemble of identical particles.
%The Pauli Exclusion Principle (PEP) is a direct implication of the Spin Statistics Theorem (SST) stated by Pauli in Ref. [1].
Despite the importance of the PEP for various areas of physics our understanding of the reasons for the PEP's existence remains unsatisfactory.
The implications of a small PEP violation on atomic transitions and on atomic properties were discussed by both Fermi and Dirac~\cite{Fer34, Dir58}.
In 1964 Messiah and Greenberg introduced a superselection rule regarding the symmetrization postulate by noting that transitions between states with different exchange symmetries are forbidden~\cite{Mes1964}. 
Moreover, in 1980 Amado and Primakoff pointed out that in the framework of quantum mechanics PEP-violating transitions are forbidden even if PEP-violation takes place~\cite{Ama80}.
As a result, PEP-violations can only be tested in "Open systems", where the testing fermion is externally introduced.

%Both Dirac and Fermi discussed the implications of a small PEP violation on atomic transitions and on atomic properties \cite{Dir58,Fer34}.
In the 1980s and 1990s several attempts were made to introduce small violations of PEP and of the spin–statistics theorem.
%In 1987 Greenberg and Mohapatra proposed that the spin–statistics theorem could have small violations.
Theoretical models implicating PEP violation were constructed by Ignatiev and Kuzmin~\cite{Ign87}, Okun~\cite{Oku87}, and Greenberg and Mohapatra~\cite{Gre87,Gre90,Moh90}. 
In these models %\cite{Ign87,Oku87,Gre87,Gre90,Moh90} 
a pair of electrons in a mixed state has the probability $\beta^2/2$ for the symmetric component and $(1-\beta^2/2)$ for the usual antisymmetric
component.
However, Govorkov showed that even a small PEP-violation leads to negative probabilities for some processes~\cite{Gov89}. 
No acceptable theoretical formalism was suggested to account for PEP violation by means of a self-consistent and non-contradictory ”small” parameter, as in the case of P- and CP-symmetry violation or L- and B-non-conservation.

In the last two decades renewed interest emerged for the development of theories with violation of statistics and for their experimental investigation.
%One of possible way to violate the PEP is to relax the basis of the Spin Statistics Theorem via violation of Lorentz symmetry at high energies.
Possible variants of PEP violation could arise from the Lorentz invariance violation at Planck energies, the existence of extra dimensions, or quantum gravity~\cite{Add2018, Add2020}.
The non-commutative quantum gravity models allow PEP violations in "Closed systems" experiments~\cite{Pis2022, Pis2023}.
%However, in scenarios involving deformations of space-time, such as in Non-Commutative Quantum Gravity , these constraints can be evaded, potentially allowing PEP violations in closed systems [Add2018]

%The experimental searches for the possible PEP violations started about 15 years later when the electron stability was tested. 
Pioneering searches for the possible PEP violations were performed by Reines and Sobel by searching for X-rays emitted in the transition of an L-shell electron to the filled K-shell in an atom~\cite{Rei74}, and by Logan and Ljubicic, who searched for $\gamma$-quanta emitted in a PEP-forbidden transition of nucleons in nuclei~\cite{Log79}.

The results of experiments are presented as lifetime limits or as limits on the relative strength of the normal and PEP-forbidden transitions.
%Critical studies of the possible violation of PEP have been done both theoretically and experimentally in \cite{Oku89,Oku89A,Ign05}. 
%More reviews and references can be found in~\cite{Hil00,SSC08}.
There are two (or four, if we consider electrons and nucleons separately) types of experiments to look for PEP violation. 
The first one is based on the search for atoms or nuclei in a non-Paulian state; the second one is based on the search for the prompt radiation accompanying non-Paulian transitions of electrons or nucleons.

Experiments searching for PEP-forbidden states were performed by Novikov~et~al.~\cite{Nov89,Nov90} and Nolte et al.~\cite{Nol91} who looked for non-Paulian exotic atoms of $^{20}$Ne and $^{36}$Ar with 3 electrons on K-shell using mass spectroscopy on fluorine and chlorine samples.
Atoms of Be with four electrons in the 1s state, which would resemble He
atoms were searched for by Javorsek et al.~\cite{Jav00}.
The non-Paulian carbon atoms in boron samples were searched for by $\gamma$-activation analysis in~\cite{Bar98}.
The nuclei of $^{5}$Li with 3 protons in the 1S-shell was searched for using time-of-flight mass spectroscopy~\cite{Nol90}.

% The same experimental data which were used to set a limit on the lifetime of the electron can be used to test the validity of the PEP for atomic electrons, as Goldhaber first pointed out.  \cite{Rei74}. 
 The same low background detector data which were used to set a limit on the lifetime of the electron disappearance or decay in an invisible channel can be used to test the validity of the PEP for atomic electrons as Goldhaber first pointed out ~\cite{Rei74}.
From the experimental point of view, the searches for characteristic X-rays due to electron decay inside an atomic shell~\cite{Fei59, Moe65, Ste75, Kov79, Bel83, Avi86, Reu91, Eji92, Aha95, Bel96, Bel99, DAMA09} are often indistinguishable from the PEP-violating transitions.
The main difference of the forbidden transition to the filled electron shell is that the energy of the emitted X-ray quanta shifts compared to normal X-rays due to the electronic screening of the nucleus.
Of course, in accordance with Amado and Primakoff~\cite{Ama80}, these transitions do not take place even if PEP is violated.

This restriction is not valid for transitions accompanied by a change of the number of identical fermions (e.g. non-Paulian $\beta^{\pm}$-transitions) and can be evaded in composite models of the electron or models including extra dimensions or the mentioned non-commutative quantum gravity models~\cite{Gre87, Add2018, Aka92}.
Although for non-Paulian $\beta^{\pm}$-decays it may be objected that the newly created fermion does not leave the nucleus and remains inside the nuclear localized system~\cite{Ell2012}.
%This restriction can be evaded... although it may be objected that..

A number of experiments have looked for the PEP violating transition of an atomic electron or nucleon in an existing system, where the symmetry of the wave-function has already been established. 
A theoretical description of this type of PEP violation contradicts the Messiah-Greenberg superselection rule or requires the use of extra-dimensions, electronic substructure, or other exotic physics~\cite{Aka92,Gre1989}.

%The Messiah–Greenberg (MG) super-selection rule [A.M. Messiah, O.W. Greenberg, Phys. Rev. 136, B248 (1964).] states that a system would not manifest a violation of PEP unless a new particle is introduced in the system from outside.
%The Messiah-Greenberg superselection rule forbids electron transitions between states of differing symmetry, allowing only newly created electrons, or electrons already in a symmetric state, to make a transition to a symmetric final state. Maijorana
%Messiah and Greenberg described a superselection rule regarding the symmetrization postulate (SP) in 1964 by noting “In summary, for systems with a fixed number of particles, there is a superselection rule between symmetry types which permits one to insert SP in the quantum theory in a consistent way.However the postulate does not appear as a necessary feature of the QM description of nature”. 
%The paper by Amado and Primakoff used different phrasing stating “Even if some principle permitted small mixed symmetry components in wave functions that are primarily antisymmetric, and kept them small, the symmetric world Hamiltonian would only connect mixed symmetry states to mixed symmetry states, just as it connects only antisymmetric states to antisymmetric states”. 

The new method was realized by Ramberg and Snow, who looked for anomalous X-rays emitted by Cu atoms in a conductor~\cite{Ram90}. 
The established upper limit on a probability for a 'new' electron passing in the conductor to form a mixed symmetry state namely non-Paulian atom with 3 electrons in the K-shell is $\beta^2/2 \leq 1.7\times10^{-26}$. 
The parameter $\beta$ corresponds to the admixed non-fermionic symmetric component of single electron level allowing the transitions to the occupied states and can be related to the ratio $\delta$ of the probabilities of PEP-forbidden and ordinary transitions $\delta^2 = \beta^2/2$~\cite{Ign87} -~\cite{Moh90}. %\cite{Ign87,Oku87,Gre87,Gre90,Moh90} 

An improvement in sensitivity has been obtained by the VIP and VIP-2 collaborations which used a copper target to bring new electrons into the system and, hence, fulfill the requirements of the super-selection rule~\cite{Bar2006, Bar2009, Cur2011, Shi2018, Pao2022, Nap2022, Por2024}. 
The obtained upper limit on the probability $\beta^2/2$ of PEP violation for traditional electrons-atoms scatterings is $\beta^2/2 \leq \mathrm{8.5\times 10^{-31}}$ and for electron diffusion model with encounters is $\beta^2/2 \leq \mathrm{2.47\times 10^{-43}}$ for 90\% C.L.~\cite{Mil2018, Por2024, Por2025}.
The upgraded VIP-3 setup will expand the number of elements being studied to higher-Z elements such as zirconium, silver, palladium, and tin~\cite{Man2024}.

Three different types of interactions between a system of fermions and a single fermion were considered by Elliott et al.~\cite{Ell2012}.
Additionally, in that work, a Pb conductor was tested instead of Cu with HPGe-detectors for dark matter and double $\beta$-decay searches. 
Lead produces higher-energy X-rays, which are less attenuated by self-shielding, and populate spectra in a region of lower relative background. 
The limit on the PEP-forbidden interactions between a system of fermions and a new previously non-interacting fermion was obtained: $\beta^2/2 \leq \mathrm{2.6\times 10^{-39}}$ for 90\% C.L.

Analysis of data obtained during the VIP-Lead experiment using "Roman lead" target led to a significant improvement in the upper bounds on the PEP-violation probabilities for electrons, when the interactions with atoms are described in terms of scatterings or encounters $\beta^2/2 \leq \mathrm{1.5\times (10^{-40}-10^{-43})}$~\cite{Pis2020}.

The Majorana low-background HPGe detector (MALBEK) at the Kimballton Underground Research Facility was used to search for PEP-violating $K_\alpha$ electron transitions using 89.5 kg-d of data. 
A lower limit on the transition lifetime of $5.8\times10^{30}$ s at 90\% C.L. was set by looking for a peak at 10.6 keV resulting from the X-ray and Auger electrons present following the transition~\cite{Abg2016}. 
The obtained lifetime limit corresponds to the constraint $\beta^2/2 \leq \mathrm{2.92\times 10^{-47}}$.

The Majorana collaboration searched for two types (I and III, in the classification in~\cite{Ell2012}) of PEP-violation using a 37.5 kg yr statistic for HPGe-detectors and set limits on the probability of an electron to be found in a symmetric quantum state~\cite{Abg2017,Arn2024}.
The most stringent upper limit on $\beta^2/2 \leq 1.03\times 10^{-48}$ was set for the type III PEP-forbidden transition of an L-shell electron in a Ge atom to the already occupied K-shell.
%The Majorana collaboration considered three models for PEP violation processes, which would produce detectable signal in the high-purity germanium detectors, and obtained limit on the PEP violating transition at $\beta^2/2 \leq 1.03\times 10^{-48}$  \cite{Arn2024}. 

The low-background Gator setup, with high-purity germanium detector, was operated at the Laboratori Nazionali del Gran Sasso, aimed at testing PEP-violating atomic transitions in lead. 
The experimental technique is based on forming a new symmetry state by introducing electrons into the existing electron system through a direct current~\cite{Bau2024}.
The obtained upper limit $\beta^2/2 \leq 4.8\times10^{-29}$ ~(90\%~C.L.) improves the previous constraint from a comparable measurement by more than one order of magnitude.

Laser atomic and molecular spectroscopy were used to search for anomalous PEP- forbidden spectral lines of $^4$He atoms~\cite{Dei95} and molecules of O$_2$~\cite{Hil96, Ang96}, and CO$_2$~\cite{Mod98}.

The violation of PEP in the nucleon system has been studied by searching for the non-Paulian transitions with $\gamma$-emission (Kamiokande ~\cite{KAMIOKANDE}, NEMO-II ~\cite{NEMO}), $p$-emission (Elegant-V ~\cite{Ejiri}, DAMA/LIBRA ~\cite{DAMA,DAMA09}) and $n$-emission~\cite{Kishimoto}, non-Paulian $\beta^+$- and $\beta^-$- decays (LSD, Kekez et al.~\cite{Kek90}, NEMO-II~\cite{NEMO}), and in nuclear $(p,p)$, $(p,\alpha)$- reactions on $^{12}$C~\cite{pp-reaction,Cur2024}.

%The Majorana collaboration considered three models for PEP violation processes, which would produce detectable signal in the high-purity germanium detectors, and obtained limit on the PEP violating transition at $\beta^2/2 \leq 1.03\times 10^{-48}$  \cite{Arn2024}. 
%The PANTHEON project aims at setting the strongest bounds on the PEP violation in nuclear reactions, focusing on the exoergic inelastic $(p,p′)$ scattering on carbon.
%The project aims at using the proton beam from the 3.5 MV Singletron accelerator of the Bellotti facility[Bellotti Ion Beam Facility] located at the Gran Sasso Underground Laboratory, to perform a measurement based on a test setup able to disentangle protons coming from the PEP-prohibited processes. They plan to improve results of \cite{pp-reaction}.  
%C. Curceanu, F. Napolitano, M. Bazzia, I. Bolognino, Acta Physica Polonica B Proceedings Supplement 17, 1-A6 (2024)
%I.J. Arnquist, F.T. Avignone III, A.S. Barabash, C.J. Bart,Majorana Demonstrator, Nat. Phys. 20, 1078-3038, 2024, arXiv:2203.02033v3

Theoretical and experimental issues in searching for PEP violation are discussed in a number of reviews %\cite{Ell2012,Mil2018,Por2024},~\cite{Oku89} -~\cite{Kap20}. 
\cite{Ell2012, Mil2018, Por2024, Oku89, Oku89A, Ign05, Bor04, Bor10, Der10, Bar10, Kap20}.
%Derbin, A.V., Fomenko, K.A. & On behalf of the Borexino Collaboration. New experimental limits on the probabilities of pauli-forbidden transitions in the 12C nucleus from data obtained with the borexino detector. Phys. Atom. Nuclei 73, 2064–2073 (2010). https://doi.org/10.1134/S1063778810120112
%Barabash, A.S. Experimental test of the Pauli Exclusion Principle. Found. Phys. 2010, 40, 703–718
%\cite{} I.G. Kaplan, Symmetry 2020, 12, 320; doi:10.3390/sym12020320
% 5. S.R. Elliott et al., Found. Phys. 42, 1015–1030 (2012)

The strongest limits for non-Paulian transitions in $^{12}$C nucleus with $\gamma$-,~$p$-,~$n$-,~$\alpha$-, and $\beta^{\pm}$- emissions were obtained with the previous Borexino detector data ~\cite{Bor04,Bor10}. 
In this paper, we present the new results obtained with the complete Borexino dataset, acquired over 14 years of lifetime. 
The large Borexino mass, its extremely low background level, and long measurement time allow us to improve the sensitivity for non-Paulian transitions with respect to previous measurements.

%Search for Pauli Exclusion Principle violations with Gator at LNGS L. Baudis1 , R. Biondi , A. Bismark, A. Clozza et al., Eur. Phys. J. C (2024) 84:1137 https://doi.org/10.1140/epjc/s10052-024-13510-1 We report on a dedicated measurement with Gator, a low-background, high-purity germanium detector operated at the Laboratori Nazionali del Gran Sasso, aimed at testing PEP-violating atomic transitions in lead.

% VIP-3 S. Manti, M. Bazzi, N. Bortolotti, C. Capoccia, Entropy 2024, 26, 752. https://doi.org/10.3390/e26090752

% reviews An Improved Limit on Pauli-Exclusion-Principle Forbidden Atomic Transitions S.R. Elliott, B.H. LaRoque, V.M. Gehman, M.F. Kidd, M. Chen, Found.Phys. 42 (2012) 1015-1030, arXiv:1107.3118, DOI: 10.1007/s10701-012-9643-y

\section{Experimental set-up and measurements}

\subsection {Brief description of Borexino} %NEW

The Borexino experiment, situated at the Laboratori Nazionali del Gran Sasso, was a large-volume liquid scintillator detector designed for the real-time spectroscopic measurement of low-energy solar neutrinos.
Its primary detection channel was through elastic neutrino-electron scattering~$(\nu, e^-)$ within a medium of exceptional radiopurity.
This combination of ultra-low background levels and a substantial target mass has also enabled us to pursue a broad physics program addressing open questions in astrophysics and particle physics.
A comprehensive description of the detector's design and components is available in previous publications~\cite{Ali02, Arp08, Arp08A, Ali08, Back12, Ito23, Ago20A, Ago18, Ago20, Bas24}.

The active core of the detector consisted of $278$~tons of liquid scintillator, composed of pseudocumene (1,2,4-trimethylbenzene, ${\mathrm{C_9H_{12}}}$) as the solvent and $1.5~$g/liter of PPO (2,5-diphenyloxazole, ${\mathrm{C_{15}H_{11}NO}}$) as the fluor.
This scintillating mixture was contained within a transparent, thin nylon Inner Vessel (IV). 
Surrounding the IV were two concentric buffer shells of pseudocumene ($323$ and $567$~tons), each doped with a light quencher (dimethyl phthalate, DMP) to suppress scintillation.
A secondary nylon membrane separated these two buffer regions, serving as a critical barrier against the inward migration of radon emanated from the outer detector components with thermal fluxes.

The entire inner detector system was housed within a spherical stainless steel sphere (SSS) with a diameter of $13.7$~meters.
This SSS was itself enclosed within a domed Water Tank (WT), measuring $18.0$~meters in diameter and $16.9$~meters in height, which held approximately $2100$~tons of high-purity water.
This water shield served as a passive barrier against external neutron and $\gamma$-radiation.

Scintillation light was detected by an array of $2212$ 8-inch PMTs uniformly mounted on the inner surface of the SSS.
All materials in contact with the scintillator or buffers were carefully selected and screened to meet stringent radiopurity requirements.
The surrounding water tank was instrumented with $208$ additional PMTs and worked as a Cherenkov muon veto (Outer Detector) to identify and tag the passage of cosmic muons.
Borexino collected data from May 2007 to October 2021~\cite{Bel25}.

\subsection{Detector calibration. Energy and spatial resolutions.}
%%----------------------------------------------------------------------------------------------
In Borexino charged particles were detected via scintillation light produced by their interactions with the liquid scintillator. 
%In principle, the energy deposited by a particle interacting in the Borexino scintillator is proportional to the amount of photons collected by the PMTs. 
The energy of an event was measured using the total collected light from all PMTs.
An electron of 1~MeV produced approximately 500 photoelectrons in the detector.
%In a simple approach, the response of the detector is assumed to be linear with respect to the energy released in the scintillator.
%The coefficient linking the event energy and the total collected charge is called the light yield (or photoelectron yield).
%Deviations from linearity at low energy can be taken into account by the ionization deficit function $ f(k_{B},E) $, where $k_B$ is the empirical Birks' constant \cite{Bir51}.
\begin{figure}
\includegraphics[bb = 30 90 500 760, width=8cm,height=10cm]{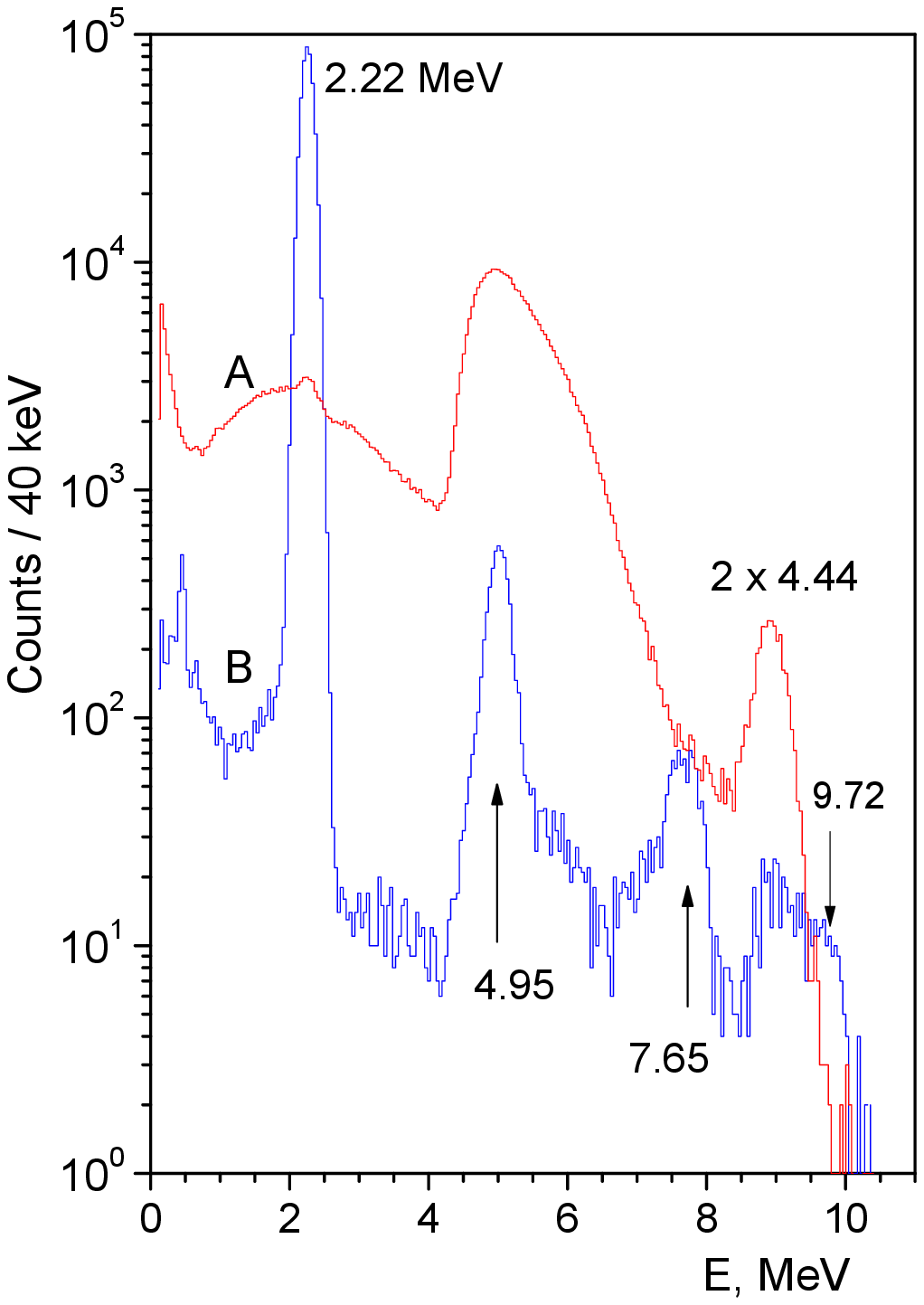}
\caption { The energy spectra of prompt (A) and delayed (B) signals registered with an ${\mathrm{^{241}Am^9Be}}$ source. The numbers indicate the energies (in MeV) of the peaks used in calibration.} %In insert, the $\gamma$-lines from neutron captures on stainless steel holder of AmBe-source are shown. } 
\label{Fig1}
\end{figure}
%\begin{figure}
%\includegraphics[bb = 30 90 500 760, width=8cm,height=10cm]{fig2_10.eps}
%\caption { The dependency of registered charge vs energy of $\gamma$-quanta (squares, left scale). The corresponding energy resolution ($\sigma_E$) is indicated on the right scale (cycles). } \label{Calib}
%\end{figure}

The detector's energy and spatial resolution were studied with radioactive sources placed at different positions inside the inner vessel.
For relatively high energies ($>$2 MeV), which are of interest for non-Paulian transition studies, the energy calibration was performed with an ${\mathrm{^{241}Am^9Be}}$ neutron source.
Neutrons were produced in two main reactions ${^9\mathrm{Be}}(\alpha,n){^{12}\mathrm{C_{gs}}}$ and $^9\mathrm{Be}(\alpha,\mathrm{n})^{12}\mathrm{C^*}$ with neutron energies up to 11~MeV and 6.5~MeV, respectively. 
The second reaction also produces one or two 4.4~MeV $\gamma$-quanta from the $\mathrm{^{12}C^*}$ de-excitation~\cite{Back12, Ito23}. 
These $\gamma$-quanta, together with the recoil protons from neutron scattering, are responsible for a prompt scintillator signal. 
Afterwards, neutrons thermalize in the hydrogen-rich organic liquid and are captured either on protons or carbon nuclei of the scintillator emitting characteristic 2.22~MeV and 4.95~MeV rays, respectively. 

In addition, neutrons are captured on iron, nickel and chromium nuclei of the stainless steel source insertion arm with the emission of $\gamma$-quanta of energies up to 9.7 MeV. 
These characteristic $\gamma$-lines produce a delayed signal in the scintillator according to the neutron capture time of $\mathrm{(254.5\pm1.8)~\mu s}$ in pseudocumene~\cite{Ago20A}. 
Figure \ref{Fig1} shows the energy spectrum of both the prompt and delayed signals originating from the $\mathrm{^{241}Am~^9Be}$ source placed at the center.
%The reactions ${^9\mathrm{Be}}(\alpha,n){^{12}\mathrm{C_{gs}}}$ and $^9\mathrm{Be}(\alpha,\mathrm{n})^{12}\mathrm{C^*}$ $\mathrm{(\gamma~4.44~MeV)}$ produce two main neutron groups with energies up to 11 MeV and 6.5 MeV, respectively.
%The resulting neutrons are thermalized by elastic and inelastic scattering in the hydrogen-rich organic scintillator and eventually are captured by protons or carbon nuclei.
%Fig.~\ref{Fig1} shows the spectrum obtained with the source placed at the center of the detector.
The upper (A, red) spectrum corresponds to the prompt neutrons and $\gamma$ rays, while the lower (B, blue) one is that of the delayed signals.

The energy scale was determined with the $2.22$~MeV and $4.95$~MeV $\gamma$ de-excitations following neutron capture on $^1$H and $^{12}$C nuclei, and with the $8.88$~MeV peak, sum of two $4.44$~MeV $\gamma$-quanta.
The expected shift of the $8.88$~MeV peak position (due to residual energy of the scattered neutron) is suppressed by the sizable quenching factor of low-energy protons.
The $7.65$~MeV $\gamma$-line following neutron capture on $^{56}$Fe present in the source holder was also used.
The deviations from linearity of the $\gamma$-peak positions were less than $30$~keV over the whole range.
Monte Carlo simulation was used to determine the energy calibration at energies above 10 MeV, since measurements with an $^{241}$Am-$^9$Be source producing $\gamma$-quanta up to 10 MeV (Fig.~\ref{Fig1}) showed (1-4)\% agreement with the MC results depending on the distance from the detector center.
The energy resolution scales approximately as ${(\sigma /E)}$ $\simeq(0.058+1.1\times10^{-3}E)/\sqrt{E}$, where E is given in MeV.
%The right boundary of the interval for the analysis was set to 16.8 MeV in accordance with the verified energy calibration of the data acquisition system.

The position of an event is determined using a photon time-of-flight reconstruction algorithm.
The resolution in the event position reconstruction is $(13\pm2)$~cm in the $x$ and $y$ coordinates, and $(14\pm2)$~cm in $z$, measured with the $^{214}$Bi-$^{214}$Po $\beta-\alpha$ decay sequence from ${\mathrm{^{238}U}}$ chain.
The energy and spatial calibration procedure is described in more detail in~\cite{Back12, Ago18}.

\section{Data analysis}
%%%%%%%%%%%%%%%%%%%%%%%%%%%%%%%%%%%%%%%%%%%%%%%%%%%%%%%%%%%%%%%%%%%%%%%%%%%%%%%%%%%%%%%%%%%%%%%
\subsection{Theoretical considerations}
\begin{figure}
\includegraphics[bb = 30 90 500 760, width=8cm,height=8cm]{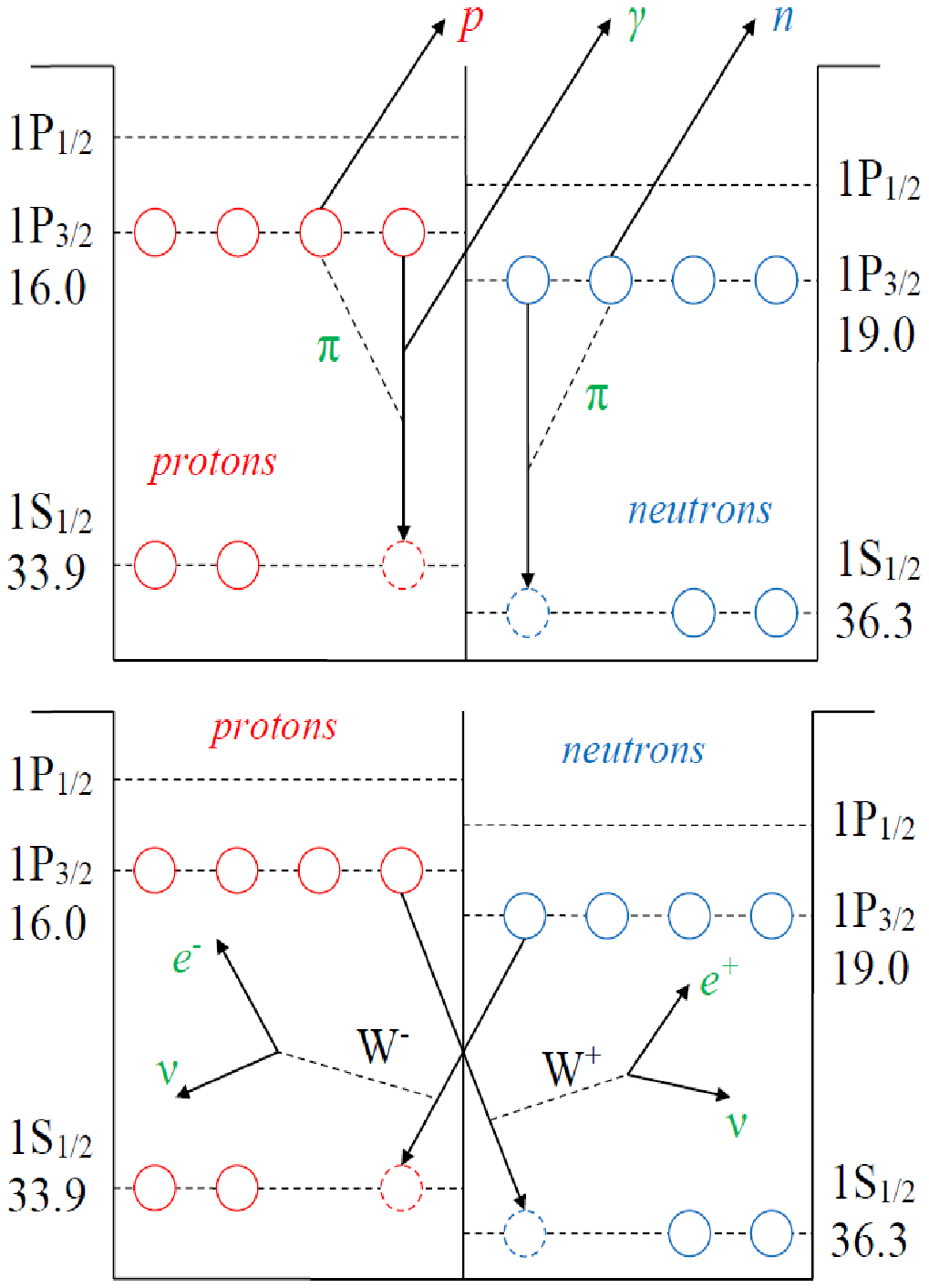}%{fig3_10.eps}
\caption {Occupation of energy levels by protons and neutrons for the $^{12}$C ground state in a simple shell model. Schemes of non-Paulian transitions of nucleons from the $P$-shell to the filled $S$-shell. Top: with $\gamma$, $n$, and $p$ emission. Bottom: with $\beta^-$ and $\beta^+$ emission. The numerical values of the energy level are given in MeV units.} \label{Fig2}
\end{figure}

%%---------------------------------------------------------------------------------------------
%The non-Paulian transitions were searched for in $^{12}$C  nuclei of the PC. 
The nucleon level scheme of $^{12}$C in a simple shell model and the hypothetical non-Paulian transitions are illustrated in Fig.~\ref{Fig2}.
%The non-Paulian transitions which have been searched for in the analysis described in this paper are schematically illustrated. 
The transition of a nucleon from the $P$-shell to the filled $S$-shell would result in an excited non-Paulian nucleus, $^{12}\widetilde{C}$.
The excitation energy corresponds to the difference between the binding energies of nucleons on $S$ and $P$ shells and is comparable with the separation energies of protons $S_p$ or neutrons $S_n$. % and $\alpha$-particles $S_{\alpha}$.
Hence, together with the emission of $\gamma$-quanta, the emission of $n$ and $p$ is possible. % and $\alpha$ is possible. 
Other reactions of interest involve weak processes violating PEP, like ${\beta}^{+}$ and ${\beta}^{-}$ decay to a non-Paulian nucleon in the final $1S_{1/2}$-state (Fig.\ref{Fig2}, bottom).

In general, the energy released in the non-Paulian transitions %under consideration 
is the difference between the binding energies of the final and initial nuclei:
\begin{center}
$Q(^{12}\mathrm{C}\rightarrow \widetilde{X}+Y)=M(^{12}\mathrm{C})-M(\widetilde{X})-M(Y)=$
\end{center}
\begin{equation}
-E_{b}(^{12}\mathrm{C})+E_{b}(\widetilde{X})+E_{b}(Y);
\end{equation}
where $\widetilde{X}$ denotes a non-Paulian nucleus, Y = $\gamma, p, n, (d, t, \mathrm{^3He}, \alpha)$ is the particle or nucleus emitted, and $E_b$ are the corresponding binding energies which are well known for ordinary nuclei~\cite{Aud95}. 
The signature of transitions, with two particles in the final state, is a peak in the measured spectrum with the width defined by the energy resolution of the detector.

In the case of non-Paulian transitions induced by weak interactions, the $\beta^{\pm}$-spectra have to be observed. 
The end-point energy of the $\beta^-$-spectrum in the reaction $\mathrm{^{12}C}\rightarrow \mathrm{^{12}\widetilde{N}}+e^{-}+\overline{\nu}_e$ is
\begin{equation}
Q=m_{n}-m_{p}-m_{e}-E_{b}(^{12}\mathrm{C})+E_{b}(^{12}\widetilde{\mathrm{N}}).
\end{equation}
A similar equation can be written for non-Paulian transition with $\beta^+$-emission $\mathrm{^{12}C}\rightarrow \mathrm{^{12}\widetilde{B}}+e^{+}+{\nu} $, but the detected energy will be shifted by $\approx2m_e$ due to the two annihilation quanta
\begin{equation}
Q=m_{p}-m_{n}-2m_{e}-E_{b}(^{12}\mathrm{C})+E_{b}(^{12}\widetilde{\mathrm{B}}).
\end{equation}

The binding energy of the non-Paulian nuclei with three neutrons or three protons on the $1S_{1/2}$-shell $E_{b}(\widetilde{X})$ can be evaluated considering the binding energy of normal nuclei $E_{b}(X)$ and the difference between the binding energies of nucleons on the
$1S_{1/2}$-shell $ E_{n,p}(S_{1/2}) $ and the binding energy of the last nucleon $ S_{n,p}(X) $:
\begin{equation}
E_{b}(\widetilde{X}_{n,p})\simeq E_{b}(X)+E_{n,p}(1S_{1/2})-S_{n,p}(X).
\end{equation}

The nucleon binding energies for light nuclei (${\mathrm{^{12}C}}$, ${\mathrm{^{11}B}}$ and others) were measured while studying $(p,2p)$ and $(p,np)$ proton scattering reactions~\cite{PNPI}. 
% with 1 GeV energy at PNPI proton synchrocyclotron
Using these data we calculated the Q-values (with uncertainties) for different non-Paulian transitions which are shown in Table~\ref{Tab1}. 
The details of the calculations can be found in our previous works~\cite{Bor04,Bor10}.

%%%%%%%%%%%%%%%%%%%%%%%% TABLE 1 %%%%%%%%%%%%%%%%%%%%%%%%%
\begin{table}[!htbp]
\begin{center}
\caption{The energies released in the transitions with non-Paulian nuclei with three neutrons or three protons on the S-shell in the final state.}\label{Tab1}
\begin{tabular}{|c|c|c|}
 \hline
 Channel & $Q_3p,~$(MeV) & $Q_3n,~$(MeV) \\ \hline
 $^{12}\mathrm{C}\rightarrow{^{12}\widetilde{\mathrm{C}}}+\gamma$ & $17.9\pm0.9$ & $17.7\pm0.6$ \\

 $^{12}\mathrm{C}\rightarrow{^{11}\widetilde{\mathrm{B}}}+ \it{p}$ & $6.3\pm0.9 $ & $7.8\pm1.0 $ \\
 $^{12}\mathrm{C}\rightarrow{^{11}\widetilde{\mathrm{C}}}+ \it{n}$ & $6.5\pm0.9 $ & $4.5\pm0.6 $ \\
  % $^{12}\mathrm{C}\rightarrow{^{8}\widetilde{\mathrm{Be}}}+\alpha$ & $3.0\pm0.6 $    &  $2.9\pm0.9  $  \\
 $^{12}\mathrm{C}\rightarrow{^{12}\widetilde{\mathrm{N}}}+ \it{e^-}+\overline{\nu}_e$ & $18.9\pm0.9 $ & - \\
 $^{12}\mathrm{C}\rightarrow{^{12}\widetilde{\mathrm{B}}}+ \it{e^+}+\nu_{e}$ & - & $17.8\pm0.9 $ \\
 \hline
 \end{tabular}
\end{center}
\end{table}

The energy released in the $\alpha$-emission reaction $^{12}\mathrm{C}\rightarrow{^{8}\widetilde{\mathrm{Be}}}+\alpha$ is $Q \cong 3.0\pm 0.6$~MeV.
As a result, the $\alpha$-particles from the decay can be found in the energy interval (1.0 - 3.0)~MeV with 90\% probability.
The light yield for an $\alpha$ with these energies corresponds to that of an electron in the energy (70 - 250) keV range.
The dominant part of the background in this range is the $\beta$-activity of $\mathrm{^{14}C}$.
The lower bound of 70~keV is close to the Borexino energy threshold, so we did not analyze this reaction using Borexino data. 
The strongest upper limit for a non-Pauli transition with $\alpha$ emission was obtained with the Borexino prototype: $\tau (\mathrm{{^{12}C} \rightarrow {^8\widetilde{Be}} + \alpha) \geq 6.1\times10^{23}}$ years (90\% C.L.)~\cite{Bor04}.
%Another reaction with a positive $Q$ value is ${^{12}\mathrm{C}}\rightarrow{^{9}\widetilde{\mathrm{B_{3p}}}}+ \mathrm{{^3H}}$ with triton emission. 
%The $Q$ value is positive only for the transition with the formation of a non-Paulian nucleus with three protons in the $1S_{1/2}$-shell.
For all other reactions such as $^{12}\mathrm{C}\rightarrow{^{10}\widetilde{\mathrm{B}}}+ d$,
$^{12}\mathrm{C}\rightarrow{^{9}\widetilde{\mathrm{B}}}+ t$, ${^{12}\mathrm{C}}\rightarrow{^{9}\widetilde{\mathrm{Be}}}+ {^3\mathrm{He}}$,
${^{12}\mathrm{C}}\rightarrow{^{6}\widetilde{\mathrm{Li}}}+ {^6\mathrm{Li}}$ and ${^{12}\mathrm{C}}\rightarrow{^{6}\widetilde{\mathrm{Li}}}+
{^4\mathrm{He}} + d$ %, except in the process ${^{12}\mathrm{C}}\rightarrow{^{9}\widetilde{\mathrm{B_{3p}}}}+ t$, 
the Q-values are negative.

%Using the obtained Q-values one can calculate the detector response for all the reactions mentioned above. 
%The recoil energy of nuclei and quenching factors for different particles have to be taken into account.
The detector response was calculated for all reactions with two- or three-particles in the final state mentioned above.
The recoil energy of nuclei and quenching factors for electrons and protons were taken into account.
%%%%%%%%%%%%%%
%Because of the uncertainties in the non-Paulian nuclei properties, the prediction of the branching ratio for the emission in each of the above mentioned channels has a poor significance. 
%For the case of the neutron disappearance (e.g. invisible decay $n\rightarrow 3\nu$) from the $1S_{1/2}$-shell in $^{12}$C nuclei, the branching ratio and spectra of the emitted particles were considered in \cite{Kam02}. 
%For the excitation energy of $^{11}$C of 17 MeV they found that the branching ratios for $p$-, $n$-, and $\alpha$-emission are of the same order of magnitude and, it is negligible for $\gamma$-emission. 
In the present paper we give the separate limits on the probabilities for each of the non-Paulian reactions using statistics obtained over 14 years of measurements. 
The obtained results are compared with the corresponding rates of normal transitions.

\subsection{Data selection and MC simulations}
%%-----------------------------------------------------------------------------------
The data selection strategy is designed to improve the signal-to-background ratio with respect to the PEP-violating process signature. 
%In this case the signature signals have to be divided into the groups of electron-like, proton-like and neutron-like signals 
The signals must be classified as electron-like, proton-like, or neutron-like and for each of the signals we develop a specific approach.

For the electron-like signal, which does not have a very specific signature
%As for the electron-like signal that does not have a very specific signatures 
with respect to the standard neutrino signal, we develop a generic procedure that would suppress the influence of cosmic muons and fast coincidence events. 
The generic candidate events are selected by the following criteria: (1) events must have a unique cluster of PMT hits; (2) events should not be flagged as muons by the outer Cherenkov detector or by the internal detector through the criterion of the hit cluster duration which is significantly longer for a non-point-like muon event; (3) events should not follow a muon within a time window of 2\,ms;
(4) events should not be followed by another event within a time window of 2\,ms except in case of neutron emission; (5) events must be reconstructed within the detector volume defined as the set of points located 0.75\,m from the inner vessel with additional parabolic cuts on the lower and upper poles.

We also apply a cosmogenic veto system:\\
 - total detector veto for $120$\,s after each muon shower identified by observation of over $20$ neutron captures in the neutron gate ($1.6$\,ms right after the muon event),\\
 - total detector veto for $4$\,s after each muon crossing the SSS,\\
 - cylindrical veto with radius of $1$\,m for $20$~s on each muon track that crosses the SSS,\\
 - spherical veto with radius of $1$\,m and duration of $20$\,s for every neutron detected in the muon gate. 

This cosmogenic veto system provides maximal background suppression at the cost of a 22.8\% exposure loss.
%This system of cosmogenic veto allows for maximal suppression of the cosmogenic background for the price of $22.8$\% exposure loss.

\begin{figure}
\includegraphics[width=1.1\columnwidth]{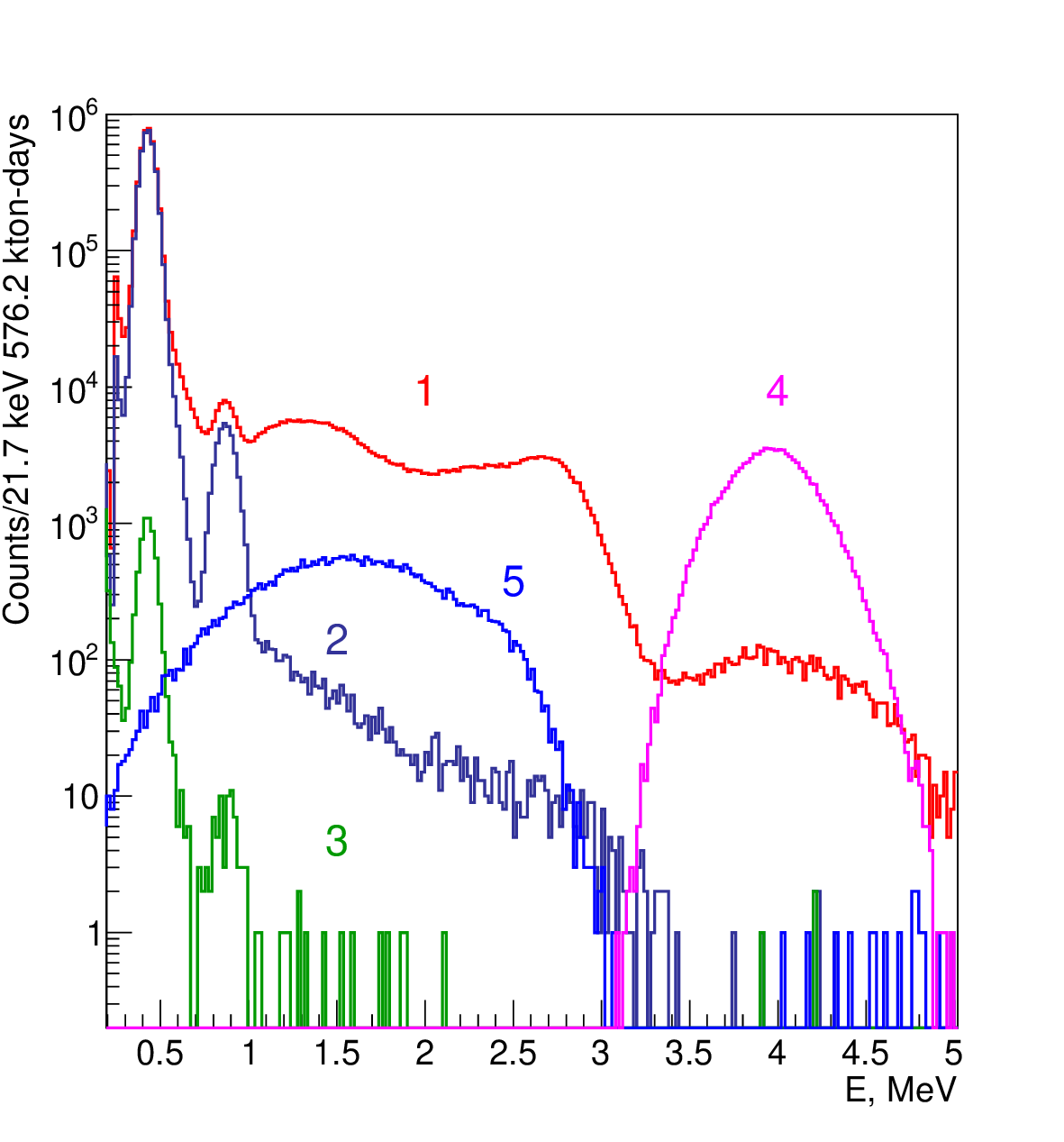}
\caption { %The fitted Borexino radial spectrum. %in terms of $R^3$ for the sake of bin volume conservation in the range $0 - 40$~m$^3$. 
 Spectra of the Borexino data and simulated PDFs corresponding to PEP-violating processes: 1~-~the generically selected data spectrum, 2~-~the spectrum of generically selected data with additional selection of proton-like events, 3~-~the spectrum of prompt proton-like events with second delayed one, 4 - simulated $\mathrm{^{12}C\rightarrow{^{11}\widetilde{\mathrm{B}}}}+ \it{p}$ decays with the same selection procedure as (2), 5 - $\mathrm{^{12}C\rightarrow{^{11}\widetilde{C}}+ \it{n}}$ decays with the same selection procedure as (3).} \label{Fig3}
\end{figure}

%Depending on the specific channel under study, pulse-shape-discrimination or delayed coincidences has also been applied to select events induced by $\gamma$, $\beta$, $p$, neutron or $\alpha$.
Proton-like events are additionally filtered using an MLP-based criterion~\cite{Bas24} originally developed for $\alpha/\beta$ discrimination %but being based mostly on the information of the hit cluster duration is quite applicable for proton discrimination as well
but being based mostly on the information of the hit cluster duration is quite applicable for proton discrimination as well (Fig.~\ref{Fig3}, line 2).
The MLP efficiency for protons is evaluated via the full MC simulation for protons since the response for PEP-forbidden processes passes exactly the same data selection procedure as the data.

Neutron-like events are selected via a delayed coincidence search. The prompt signal, arising from neutron moderation, is proton-like, while the delayed signal comes from radiative neutron capture. 
Coincident events are required to be within 1.5 m in distance and 1 ms in time (Fig.~\ref{Fig3}, line 3).
%The neutron-like events are selected by delayed coincidence search, where the prompt signal is coming from the neutron moderation that is a proton-like signal and the delayed signal is the neutron radiative capture that is searched within 1.5 in distance and 1 ms after the prompt event (Fig.~\ref{Fig3}, line 3).

\begin{figure}
\includegraphics[width=1.1\columnwidth]{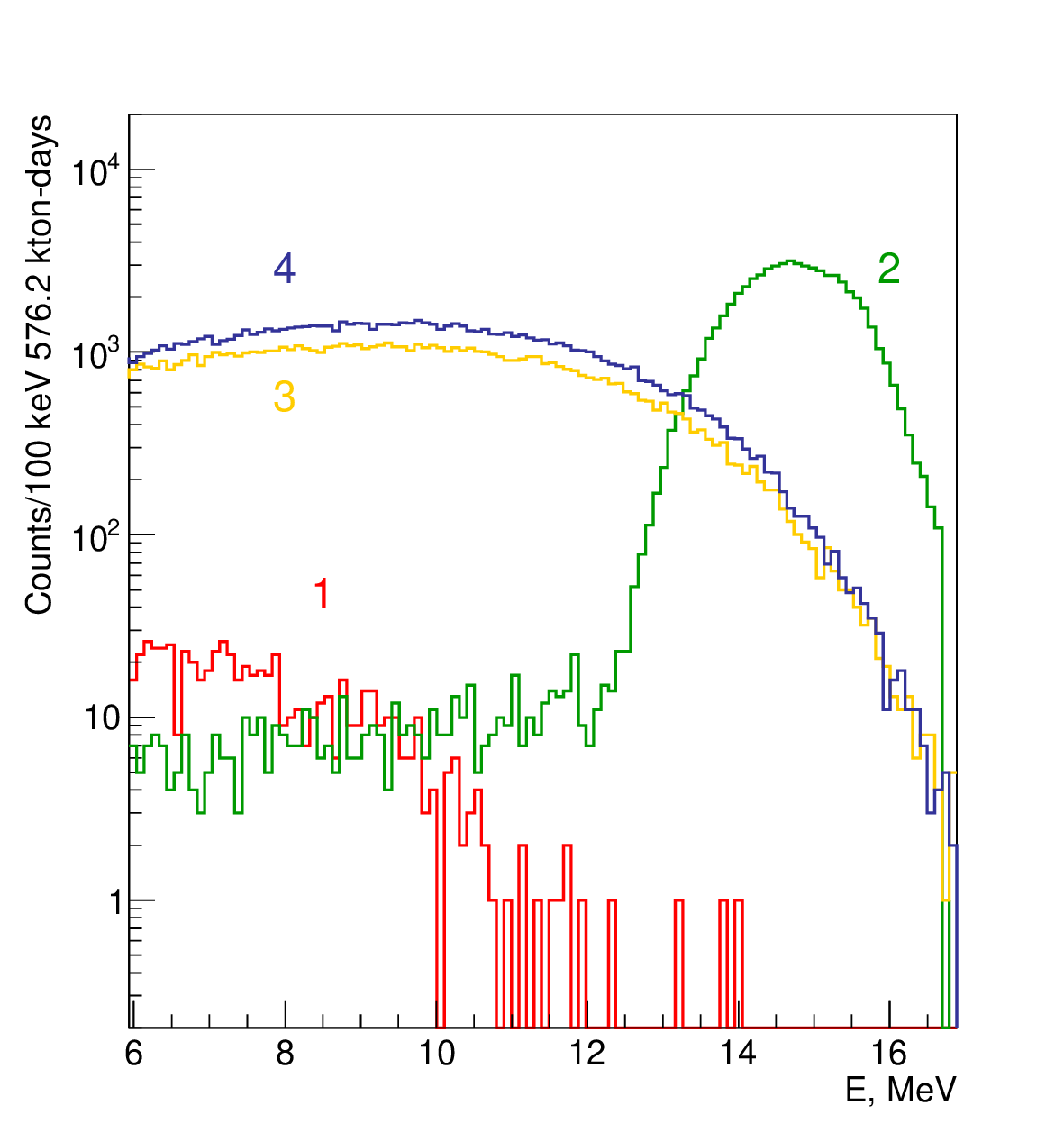}
\caption {Spectra of the Borexino data and simulated PDFs corresponding to PEP-violating processes: 1~-~the generically selected data spectrum, 2~-~simulated $\mathrm{^{12}C\rightarrow{^{12}\widetilde{\mathrm{C}}}+\gamma}$ decays, 3~-~$\mathrm{^{12}C\rightarrow{^{12}\widetilde{N}}+ e^- + \overline{\nu}_e}$, 4~-~ $\mathrm{^{12}C\rightarrow{^{12}\widetilde{B}}+ e^+ + \nu_e}$.}
\label{Fig4}
\end{figure}

The spectra of the Borexino data passing the data selection procedure are shown in Figs.~\ref{Fig3} and~\ref{Fig4}.
The Borexino spectrum in Fig.~\ref{Fig3} is shown in the range (0.2 - 4.5)\,MeV. 
The contribution from the PEP forbidden transitions with proton or neutron emission was searched in this range. 
The spectrum in Fig.~\ref{Fig4} is shown in the range of (6.0 - 17)\,MeV which was used to search for non-Paulian transitions with $\gamma$-emission and $\beta^\pm$-decays.
%\subsection{Monte-Carlo simulations}

%The PEP-violating processes signal response was evaluated via simulation with Monte-Carlo (MC) method.
The signal response for PEP-violating processes was evaluated using Monte Carlo (MC) simulations.
The simulation was performed with the advanced MC simulation code developed for the Borexino detector solar neutrino program~\cite{Ago18}. 
The code was carefully fine-tuned on the source calibration data and reproduces both temporal and amplitude properties of the detector signal. 
Since the MC code output has exactly the same structure as the real detector data output, it was essential to perform the same data selection procedure that was used for the real detector data followed by evaluation of the selection procedure efficiency. 

In order to take into account the edge effects near the fiducial volume boundary, it was decided to originate the primary MC event in a sphere with radius of 5\,m that covers the whole fiducial volume taking into account all possible cross-propagation effects related to position reconstruction uncertainties and ionization distributions.
%cross-propagation related to both position reconstruction uncertainties and real ionisation distributions. 
The resulting energy spectra of the PEP-violating processes signal that pass the data selection procedures are shown in Fig.~\ref{Fig3} and Fig.~\ref{Fig4}. 

Table \ref{Tab2} shows the energy range $\Delta E$ in which PEP-violating decays were searched for; 
the total detection efficiency $\epsilon(\Delta E)$ defined by MC simulations that take into account the fiducial volume, the energy interval $\Delta E$, and the efficiency of all cuts applied to real data; 
the observed number of events $N_{obs}$ in the range and the expected background $N_{exp}$. 
The right column shows the upper limit on the number of candidate events $S_{lim}$ registered in the $\Delta E$ energy interval for 90~\% confidence level and calculated according to the Feldman-Cousins procedure~\cite{Fel98}. 
The number of expected background events $N_{exp}$ is set to the minimum number of events that we know for sure existed, so we take a noticeably conservative approach.

%The choice of studied ranges $(3.5-5.0)$~MeV and $(2.0-3.0)$~MeV  for processes with proton and neutron emission is determined by the requirement of high registration efficiencies, for higher-energy processes $(\geq~ 12.5$~MeV) with $\gamma$- and $\beta^\pm$-emission - by the ratio of the expected effect to the background.

The (1.8\,-\,5.2)\, MeV and (2.2\,-\,3.7)\, MeV ranges were chosen for the proton and neutron emission channels, respectively, to ensure high detection efficiencies. For higher-energy processes $(\geq12.5$~MeV) involving $\gamma$ and $\beta^\pm$ emission, the choice was guided by the expected signal-to-background ratio.
%/////////////////////////////////////////////////////////////%
%The experimental energy spectra of Borexino in the range 1.0-14 MeV, collected during the whole detector exposure, is shown in Fig.\ref{Spectr_all}.  The raw spectrum is presented at the top. At energies below 3~MeV, the spectrum is dominated by 2.6 MeV $\gamma$'s from the $\beta$-decay of $^{208}$Tl due to radioactive contaminations in the PMTs and in the SSS.

\begin{table}[ht]
\caption{The analysis ranges $\Delta E$~(MeV), total efficiencies $\varepsilon(\Delta E)$ (\%), number of observed $N_{obs}$ and expected $N_{exp}$ events, and the upper limits on number of counts for PEP-violating reactions $S_{lim}$ (90\% C.L.). The proton emission analysis was performed by two methods (see text).}
 \centering
 \begin{tabular} {|c|c|c|c|c|c|c|c|}
 \hline
 Reaction & $\Delta E$ & $\varepsilon$ & $\!\!N_{\!obs} $ & $\!\!N_{\!exp}$ & $\!\!S_{lim}$ \\\hline
 $\mathrm{{^{12}C}\rightarrow{^{12}\widetilde{C}}}+\gamma$ & 12.5-17 & 27.0 & 3 & 2.8 & 4.62 \\
 $\mathrm{{^{12}C}\rightarrow{^{11}\widetilde{B}}}+\it{p}$ & 3.7-5.2 & 27.1 & 2 & 0 & 5.91\\
 $\mathrm{{^{12}C}\rightarrow{^{11}\widetilde{B}}}+\it{p}$ & 1.8-4.6 & 27.1 & - & - & 48.8\\
 $\mathrm{{^{12}C}\rightarrow{^{11}\widetilde{C}}}+\it{n}$ & 2.2-3.7 & 2.68 & 0 & 0 & 2.44 \\
 $\!\!\mathrm{{^{12}\!C}\!\rightarrow\!{^{12}\!\widetilde{N}}}\!+\!\it{e^-}\!\!+\!\overline{\nu}_e$ & 12.5-17 & 3.28 & 3 & 2.8 & 4.62 \\
 $\!\!\mathrm{{^{12}\!C}\!\rightarrow\!{^{12}\!\widetilde{B}}}\!+\!\it{e^+}\!\!+\!\nu_{e}$ & 12.5-17 & 3.37 & 3 & 2.8 &4.62\\\hline
 \end{tabular}
 \label{Tab2}
\end{table}

%The contribution to the background in low-energy regions where additional events are expected during PEP-violation transitions with the emission of proton or neutron can be due to many processes, firstly $\mathrm{{^{208}Tl}}$-decays and solar $\mathrm{^8B}$-neutrinos.
The background in the low-energy regions, where events from PEP-violating proton or neutron emission are expected, arises from several sources, primarily $\mathrm{{^{208}Tl}}$ decays and solar $\mathrm{^8B}$ neutrinos.
However, the additional MLP-based criteria for protons and the coincidence of two signals for neutrons make the observed background $N_{obs}$ extremely low, and the expected background $N_{exp}$ is difficult to predict.
Since the expected accuracy of such calculations is quite low, we decided to make conservative estimate of $S_{lim}$ under the assumption of zero expected background ($N_{exp}$ = 0) for two studied intervals (Tab. \ref{Tab2}, lines 3, 4).
The expected background at high energies (above 12.5~MeV) is $N_{exp}$ = 2.8 events and is associated mainly with solar $\mathrm{{^8B}}$-neutrino, atmospheric neutrinos and missed long-lived cosmogenic isotopes ($\mathrm{^8B}$, $\mathrm{^8Li}$, etc.)~\cite{Ago20}.

%A possible contribution to the background in the region above 12.5~MeV could be due, for example, to missed decays of long-lived cosmogenic isotopes ($\mathrm{^8B}$, $\mathrm{^8Li}$, etc.), solar $\mathrm{{^8B}}$-neutrino or atmospheric neutrinos \cite{Ago20}. 

\section{Results}
%%%%%%%%%%%%%%%%%%%%%%%%%%%%%%%%%%%%%%%%%%%%%%%%%%%%%%%%%%%%%%%%%%%%%%
The lower limits on the lifetime for PEP violating transitions of nucleons from $P$-shell to the occupied $1S_{1/2}$-shell were obtained using the formula:
\begin{equation}
 \tau \geq \frac{N_{N}N_{n}}{S_{lim}}T\varepsilon(\Delta E),
 \label{TLimit}
\end{equation}
where $N_{N}=2.04\times10^{31}$ is the number of $\mathrm{{^{12}C}}$ nuclei in 5~m radius PC volume, $N_{n} = 8$ (or 4 for specific channels) is the number of nucleons ($n$ and/or $p$) that could undergo non-Paulian transitions, and $T$ is the time of measurements.
% total statistic is 576.2 kt days
The values of efficiencies $\varepsilon(\Delta E)$ and upper limits on the number of candidate events registered in the $ \Delta E $ energy interval $S_{lim}$ are given in Table~\ref{Tab2}. 

The mass of scintillator inside the selected volume, which is proportional $N_N$ as well as live time of measurement $T$, are known with good accuracy, better than 0.2\%. 
The systematic uncertainties of the other two parameters, $\varepsilon(\Delta E)$ and $S_{lim}$, are significantly larger, as they depend on the poorly constrained $Q$ values of PEP-forbidden reactions.
%Specifically, the determined efficiency depends significantly on the $Q$-value. 
The increase or decrease in $Q$-value by 1 MeV for the $\beta^\pm$-decay process leads to a change in the efficiency for events with $E_e \geq 12.5$ MeV by an additional ~5\%.
The values of $\varepsilon(\Delta E)$ are calculated for the conservative $Q$ values and as a result the lifetime limits are calculated in accordance with Eq.\,(\ref{TLimit}) with $\varepsilon(\Delta E)$ and $S_{lim}$, which gave conservative limits.

\subsection{Limits on PEP-forbidden transitions with $\gamma$-emission: $\mathrm{{^{12}C}\rightarrow{^{12}\widetilde{C}}}+\gamma$ } 
%%---------------------------------------------------------------------
The limit on the probability of the non-Paulian transitions $\mathrm{^{12}C\rightarrow{^{12}\widetilde{C}}}$+$\gamma$ is based on the experimental fact that only three events above 12.5 MeV survive the selection cuts.
%As shown in Table~\ref{Tab1}, the most probable energy of $\gamma$-quanta emitted in the nucleon transition from the shell $1P_{3/2}$ to the shell $1S_{1/2}$ is $\simeq$~17.8~MeV.
%Taking into account the error of Q-values, the energy of $\gamma$-quanta is inside the energy interval (16.4$\div$19.4) MeV with 90\% probability. 
As shown in Table~\ref{Tab1}, the $\gamma$-quanta energies lie within the $(16.4-19.4)$~MeV interval with 90\% probability.
%The efficiency of $\gamma$ detection is found for the conservative value $E_{\gamma}$=16.4 MeV. 
The $\gamma$ detection efficiency was calculated using the conservative value $E_{\gamma}$=16.4 MeV.
As stated above, the uniformly distributed $\gamma$'s were simulated in 5~m radius sphere - inside the 4.25 m radius inner vessel (PC + PPO) and in the 0.75~m- thick layer of buffer (PC+DMP) surrounding the inner vessel.
The response function is shown in Fig.\,\ref{Fig4} (line 3). The efficiency for detecting 16.4~MeV $\gamma$-quanta in the $(12.5-17.0)$~MeV range is $\varepsilon(\Delta E)=0.270$ (27\%).

The number of $ ^{12}\mathrm{C}$ target nuclei in 460~tons of PC is $ N_{N}=2.04\times10^{31}$ (taking into account the isotopic abundance of $^{12}$C). 
The number of nucleons on the $P$-shell is $N_{n}=8$, the total data taking time is $T=10.94$~y, and the upper limit on the number of candidate events is $S_{lim}=4.62$ with 90\%~C.L. in accordance with the Feldman-Cousins procedure~\cite{Fel98}. 
The limit obtained using the cited numbers equals to:
\begin{equation}%\label{limgamma}
\tau_{\gamma} (\mathrm{{^{12}C}\rightarrow {^{12}\widetilde{C}}}+\gamma )\geq 1.05\times10^{32}\; \mathrm{y}, \label{GLimit}
\end{equation}
for the 90\% C.L. 

The result improves our previous limit by a factor of two~\cite{Bor10} and is identical to the limit obtained with the Kamiokande detector for $\mathrm{{^{16}O}}$ nuclei: $\tau (\mathrm{{^{16}O}\rightarrow {^{16}\widetilde{O}}}+\gamma )\geq 1.0\times 10^{32}$ y for $\gamma$'s with energies $(19-50)$ MeV ~\cite{KAMIOKANDE}. 
These limits are significantly stronger than the one obtained with the NEMO-2 double $\beta$-decay detector: $\tau (\mathrm{{^{12}C}\rightarrow {^{12}\widetilde{C}}}+\gamma )\geq 4.2\times 10^{24} $ y~\cite{NEMO}.

The limit on the total nucleon lifetime can be derived from $\tau_{\gamma}$ as $\tau = \tau_{\gamma}\mathrm{Br}(\gamma)$, where $\mathrm{Br}(\gamma) = \Gamma_{\gamma}/\Gamma_{tot}$ is the $\gamma$-decay branching fraction. 
For the case of $^{16}\mathrm{O}$ nucleus the calculated value of $\mathrm{Br}(\gamma)$ is inside interval $(2.7-10.4)\times10^{-5}$~\cite{KAMIOKANDE}.
The advantage of Borexino is the ability to detect other modes of non-Paulian transitions in nuclei with $p$-, $n$- or $\beta^\pm$-emission.
%Unlike the Kamiokande, the Borexino can directly detect the non-Paulian transitions with $p$-, $n$- or $\alpha$-emission.

\subsection{Limits on PEP-forbidden transitions with proton emission: $\mathrm{^{12}C\rightarrow {^{11}\widetilde{B}}}$ + $p$} 
%%%% PROTON--------------------------------------------------------
The energy released in these non-Paulian transitions is within the (5.0-9.0) MeV interval with a probability of 90\% (Table~\ref{Tab1}). 
Taking into account the recoil energy of $\mathrm{{^{11}\widetilde{B}}}$ nucleus, the energy of the proton is $(4.6-8.3)$~MeV.

%The response functions of protons with different energies were calculated using MC simulation which takes into account the quenching factor for protons which %The empirical Birks' constant \cite{Bir51} 
%was determined from the spectrum of recoil protons measured with $\mathrm{{^{241}Am} {^9Be}}$-source \cite{Back12}. 
The response functions for protons of different energies were calculated using MC simulations. 
These simulations account for the proton quenching factor, which was determined from the spectrum of recoil protons measured with an $\mathrm{{^{241}Am} {^9Be}}$-source~\cite{Back12}.

%It was determined that the light yield of PC for a proton with the energy $E_{p}$=4.6~(8.3)~MeV corresponds to an electron energy of $E_{e}$=1.8~(4.6)~MeV and the proton peak can be found in the energy interval $(1.8-4.6)$~MeV which is much wider than the energy resolution of the detector ($\sigma_E\cong$ 80 keV for $E_e$ = 2~MeV). 
The light yield for a proton with energy $E_{p}$\,=\,4.6\,(8.3)\,MeV corresponds to that of an electron with $E_{e}$\,=\,1.8\,(4.6)\,MeV. Consequently, the proton peak appears in the $(1.8\,-\,4.6)$\,MeV interval, which is significantly broader than the detector’s energy resolution ($\sigma_E\cong$\,80\,keV at $E_e$\,=\,2\,MeV).
%The response function for 8.3\,MeV protons is shown in Fig.\,\ref{Fig3}, line 4.

%First, we looked for the proton's peak in the spectrum of single events obtained with FV cut (line 2, Fig.\ref{Fig3}). 
The spectrum of generically selected data with additional selection of proton-like events (line 2, Fig.\,\ref{Fig3}) was fitted by polynomial function and response functions for protons with different energy positions.
%Except the region of 2.614 MeV $\gamma$ peak, 
This procedure gives maximum value $S_{lim}=48.8$ (90\%~C.L.) for protons with an initial energy 4.6 MeV. 
The lower limit on the lifetime was found from the formula (\ref{TLimit}) taking into account the corresponding cut efficiency $\epsilon=0.271$:
\begin{equation}\label{limp1}
\tau_p (\mathrm{{^{12}C}\rightarrow {^{11}\widetilde{\mathrm{B}}}}+\it{p} )\mathrm{\geq 9.96\times 10^{30}}\; \mathrm{y}\; (90\%\: C.L.).
\end{equation}
A more stringent limit is obtained for protons with an initial energy of $E_p=8.3$~MeV. The response function is shown in Fig.~\ref{Fig3}, line 4.
%Since the number of events in the analyzed $3.3\sigma$ interval $(3.7-5.2)$~MeV (Table~\ref{Tab2}) is only two and the expected background is zero, we used eq.~(\ref{TLimit}) to establish the lower limit on the lifetime for the PEP-forbidden transition with emission of 8.3 MeV proton:
Since only two events are observed in the $3.3\sigma$ interval $(3.7-5.2)$~MeV (Table~\ref{Tab2}) and the expected background is zero, we used Eq.~(\ref{TLimit}) to establish the lower limit on the lifetime for the PEP-forbidden transition with an 8.3 MeV proton:
\begin{equation}\label{limp2}
\tau_p (\mathrm{{^{12}C}\rightarrow {^{11}\widetilde{\mathrm{B}}}}+\it{p} )\mathrm{\geq 8.22\times 10^{31}}\; \mathrm{y}\; (90\%\: C.L.).
\end{equation} 
%The corresponding upper limits on nuclear instabilities of $^{12}\mathrm{C}$ nucleus differ from the limits (\ref{limp1}) and (\ref{limp}) by factor $N_n$=8.
As noted above, the estimating of the expected background in this region is quite difficult, so the conservative value of $N_{exp}=0$ was used. 
If one assumes that the expected background value corresponds to the measured value of $N_{exp}=N_{obs}=2$, then the limit (8) becomes 1.5 times more stringent.

The obtained upper limits on the nuclear instabilities of $^{12}\mathrm{C}$ are a factor of $\sim$(10-100) stronger than our previous result $\tau_p \geq 8.9\times10^{29}$~y~\cite{Bor10}.
%The main reason for the increase in sensitivity is due to the use of MLP analysis of the pulse shape, which made it possible to reduce the background level in the energy region where the proton can give a signal.
The increase in sensitivity is primarily due to the use of MLP-based pulse-shape analysis, originally designed for $^{210}$Po discrimination~\cite{Bas24}, which reduced the background level in the energy region where a proton signal is expected.
We adopt the conservative limit from Eq.~(\ref{limp1}) for further analysis.
For comparison, the result obtained with the 250\,kg NaI DAMA/LIBRA detector is $\tau(\mathrm{{^{23}Na}, {^{127}I}\rightarrow {^{22}\widetilde{Ne}},{^{127}\widetilde{Te}}}+\it{p}) \mathrm{\geq 1.9\times 10^{25} y}$ (90\% C.L.) for protons with $E_p \geq$~10~MeV~\cite{DAMA09}.

%The energy of $\alpha$-particles emitted in $\mathrm{{^{12}C}\rightarrow {^{8}\widetilde{Be}}}+\alpha$ decay can be found in the 1.0-3.0 MeV interval. Because of the quenching factor, it corresponds to an electron energy range $(70-250)$ keV. The energy 70 keV is close to the Borexino lower energy threshold and we have not analyzed this reaction with the Borexino data. Our limit on this mode of transition, which was obtained using the CTF measurements with 20 keV threshold, is $\tau(\mathrm{{^{12}C}\rightarrow {^{8}\widetilde{Be}}}+\alpha) \geq 6.1\times 10^{23}$ y (90\% C.L.).

\subsection{Limits on PEP-forbidden transition with neutron emission: $\mathrm{{^{12}C}\rightarrow {^{11}\widetilde{\mathrm{C}}}}$+$n$} %%%NEUTRON------------------------------------------------------
 The kinetic energy of the initial neutron is in the $(3.2\,-\,7.3)$ MeV interval with 90\% probability (Table \ref{Tab1}). 
The fast neutrons produced in the PEP-forbidden reaction thermalize in the $\mathrm{{C_9H_{12}}}$ organic scintillator and are then captured by protons or $\mathrm{{^{12}C}}$ nuclei with a mean capture time of $\tau_n \cong 255\,\mu\mathrm{s}$. 
The capture reaction $n+p\rightarrow d+\gamma$ is followed by the emission of a 2.2\,MeV $\gamma$-ray. 
The cross section for the capture on a proton for a thermal neutron is $\sigma _{\gamma} =330$\,mb.
For capture on $\mathrm{{^{12}C}}$ nuclei, the cross sections are much smaller ($ \sigma _{\gamma}$ = 3.5\,mb, $E_{\gamma}=4.95$\,MeV). 
As a result, the 4.95~MeV peak intensity is about 1\% of the 2.2\,MeV peak.
The response function of Borexino to the $\gamma$'s of 2.2\,MeV energy was precisely measured with ${\mathrm{{^{241}Am}{^9Be}}}$ neutron source~\cite{Back12}.

In principle, the background levels at 2.2\,MeV could be used to obtain an upper limit on the number of 2.2~MeV $\gamma$'s, and hence a limit on the probability of PEP-forbidden neutron emission. 
%neutron production in the reaction $\mathrm{{^{12}C}\rightarrow {^{11}\widetilde C}}+\it{n}$.
%The position and width of the peak is well known, the fitting procedure gives $S_{lim}$=57. 
%Using equation (\ref{TLimit}) one can obtain the limit on the probability on neutron emission: $\tau_n ({^{12}\mathrm{C}}\rightarrow {^{11}\widetilde{\mathrm{C}}}+n)\geq 8.1 \times 10^{29} \mathrm{y}$ (90\% C.L.)

Higher sensitivity can be achieved by selecting two correlated events: the first signal from recoil protons and the second, a 2.2\,MeV $\gamma$-ray from neutron capture.
Candidate events were selected among correlated event pairs occurring within 1.0~ms (4$\tau_n$) of each other, excluding coincidence times below 20\,$\mu$s. 
The prompt event must be proton-like event with an electron equivalent energy in the range $(2.2\,-\,3.7)$\,MeV (Fig.\,\ref{Fig3}, line 2). 
The expected response function of prompt event for 7.3\,MeV neutrons is shown in Fig.\,\ref{Fig3}, line\,5.

%These events are additionally filtered by the MLP-based criterion [75] that is developed forα/β discrimination but being based mostly on the information of the hit cluster duration is quite applicable for proton discrimination as well (Fig. 3, line 2).The neutron-like events are selected by delayed coincidence search, where the prompt signal is coming from the neutron moderation that is a proton-like signal and the delayed signal is the neutron radiative capture that is searched within 1.5 in distance and 1 ms after the prompt event (Fig. 3, line 3).
%The lower threshold is defined by the minimal neutron energy 3.2 MeV (visible energy of 0.6~MeV) taking into the rate of random coincidences.
%The response functions for neutrons with energy 3.0 and 6.0 MeV are shown in Fig.\ref{response}. 
The energy of the second event was required to be in $(1.0\,-\,2.4)$\,MeV interval to ensure high detection efficiency for the 2.2\,MeV $\gamma$-rays. 
Additionally, the position of the second event has to be within 1.5\,m in distance and 1\,ms after the prompt event. 
The energy spectrum of the prompt events selected with these criteria is shown in Fig.\,\ref{Fig3}, line\,3.
%If $E_n$ exceeds the energy of first exited state of $^{12}\mathrm{C}$ then the high energy part connected with detection of 4.44 MeV $\gamma$'s appears in the spectrum of the prompt events. The maximal value of the correlated events $N$=26 was found for the ranges (0.6--2.3) MeV and (4.3--5.0) MeV that corresponds 6 MeVneutrons. Taking into account the probability to find 6.0 MeV neutrons signal in these range ($\varepsilon$=0.9), efficiency of registering 2.2 MeV $\gamma$'s ($\varepsilon$=0.96), full number of $^{12}$C in the inner vessel $N_N$=1.24$\times$10$^{31}$ and $S_{lim}$=33 for 90\% C.L., the limit is:
%The procedure did not find any event satisfying the above cuts in the $(2.2-3.7)$~MeV range, that with the expected zero background, leads to the limit:
No events satisfying these criteria were found in the $(2.2-3.7)$~MeV range. With an expected background of zero, this leads to the limit:
\begin{equation}\label{limn}
\tau_n({^{12}\mathrm{C}}\rightarrow {^{11}\widetilde{\mathrm{C}}}+n) \geq 1.97\times 10^{31} \mathrm{y}~(90\%~C.L.).
\end{equation}

The resulting limit improves our previous result by a factor of 8~\cite{Bor10} and is 9 orders of magnitude stronger than the one obtained through searching for spontaneous neutron emission from lead: $\tau(\mathrm{Pb}\rightarrow \widetilde{\mathrm{Pb}}+n)\geq 2.1\times 10^{22}\; y\: (68\%\: C.L.) $~\cite{Kishimoto}.

\subsection{Limits on PEP-forbidden $\beta^{\pm}$-transitions: {${^{12}\mathrm{C}}\rightarrow {^{12}\widetilde{\mathrm{N}}}+e^{-}+\overline{\nu}_e$} and
$ {^{12}\mathrm{C}}\rightarrow {^{12}\widetilde{\mathrm{B}}}+e^{+}+\nu_e $} 
%%%%%BETA-----------------------------------------------------------
The high-energy portion of the Borexino spectrum for events selected by the generic criteria is shown in Fig.\,\ref{Fig4}, line 1.
The energy released in the reaction $\mathrm{{^{12}C}\rightarrow {^{12}\widetilde{N}}}+e^{-}+\overline{\nu}_e $ is $(18.9\pm0.9)$\,MeV. 
Since the interval is relatively narrow, the analysis was performed using the most probable Q-value.
 The shape of the $\beta^{-}$ spectrum, calculated by MC method with the most probable endpoint energy of 18.9~MeV, is shown in Fig.~\ref{Fig4}, line 2. 
%The spectrum was determined by MC method. 
The limit on the non-Paulian $\beta^-$-transition probability is based on the observation of only three events with $E_e\geq12.5$~MeV. % not accompanied by a muon veto signal.
The MC-derived detection efficiency for electrons with $E_{e}>12.5 $~MeV is $\varepsilon=0.0328$. %0.1027 is part of 18.9 beta-spectrum above 12.5 MeV
The lower limit on the lifetime of the $^{12}$C nucleus for the PEP-violating transition involving 4 neutrons ($N_n$=4) is then:
\begin{equation}\label{limbeta-}
\tau_{\beta^-} ({^{12}\mathrm{C}}\rightarrow {^{12}\widetilde{\mathrm{N}}}+e^{-}+\overline{\nu}_e)\geq 6.39\times 10^{30}\: \mathrm{y}\:
(90\%~C.L.).
\end{equation}
This result is more than six orders of magnitude stronger than that obtained by NEMO-2, $\tau({^{12}\mathrm{C}}\rightarrow
{^{12}\widetilde{\mathrm{N}}}+e^{-}+\overline{\nu}_e )\geq 3.1\times 10^{24}\: \mathrm{y}\: (90\%\:~C.L.)$~\cite{NEMO}.

Data from the LSD detector~\cite{LSD}, located in the tunnel under Mont Blanc, allow a qualitative limit to be set for this decay mode. 
In Ref.~~\cite{Kek90}, it is reported that only two events were observed with energies higher than 12~MeV during 75~days of data taking with the detector loaded with 7.2~tons of scintillator, containing $ 3\times 10^{29} $ $^{12}$C nuclei.
%The upper limit that can be obtained using formula (\ref{TLimit}) with these data (with $S_{lim}$=5.91 events for 90\%~C.L. and detection efficiency ${\varepsilon (\mathrm{E}\geq \mathrm{12~MeV})} =0.23 $ is $\tau (^{12}$C$\rightarrow^{12}\widetilde{\mathrm{N}}+e^{-}+\overline{\nu}_e)\geq  9.5\times10^{27}$ y (90\%~C.L.).
Using Eq.~(\ref{TLimit}) with these data ($S_{lim}$=5.91 events at 90\% C.L.) and detection efficiency ${\varepsilon (\mathrm{E}\geq \mathrm{12~MeV})} =0.23 $ yields an upper limit of $\tau ({^{12}\mathrm{C}}\rightarrow{^{12}\widetilde{\mathrm{N}}}+e^{-}+\overline{\nu}_e)\geq 9.5\times10^{27}$~y (90\%~C.L.).

The expected endpoint energy of the $\beta^+$ spectrum is 17.8~MeV, but the detected spectrum is shifted upward by $\simeq0.85 $~MeV due to the presence of annihilation quanta (Fig.~\ref{Fig4}, line 4). 
The efficiency of the $^{12}$C$\rightarrow {^{12}\widetilde{\mathrm{B}}}+e^{+}+\nu_e$ transition detection with energy release $ E>12.5 $\,MeV is $\varepsilon =0.0337$. % 0.1079 is part of 17.8 beta-plus spectrum above 12.5 MeV
The lower limit on the lifetime of the $^{12}$C nucleus is then
\begin{equation}\label{limbeta+}
\tau_{\beta^+} ({^{12}\mathrm{C}}\rightarrow {^{12}\widetilde{\mathrm{B}}}+e^{+}+\nu_e )\geq 6.56\times 10^{30}\: \mathrm{y}\: (90\%\: C.L.)
\end{equation}
The limits obtained by the NEMO-2 collaboration for this reaction are more than 6 orders of magnitude weaker: $ \tau
(^{12}\mathrm{C}\rightarrow ^{12}\widetilde{\mathrm{B}}+e^{+}+\nu_e)\geq 2.6\times 10^{24}\: y\: (90\%\: C.L.) $~\cite{NEMO}.
The final limits on the nucleon instability are shown in Table~\ref{Lim} in comparison with the previous results obtained for the same PEP-violating transitions. 
The limit~\cite{DAMA09} relates to the instability of $^{23}\mathrm{Na}$ and $^{127}\mathrm{I}$ nuclei, all other limits are given per nucleon for which the non-Paulian transition is possible.

\begin{table}[!htbp]
\begin{center}
\caption{Mean lifetime limits for non-Paulian transitions of nucleons in the Borexino in comparison with previous results.} \label{Lim}
\begin{tabular}{|l|c|c|c|}
 \hline
 Channel & $\tau_{lim}$ (y) & Previous\\
 & 90\% C.L. & non-Borexino limits \\ \hline
 $\mathrm{{^{12}C}\rightarrow{^{12}\widetilde{C}}}+\gamma$ & 1.1$\times10^{32}$ & 4.2$\times10^{24}$($\mathrm{{^{12}C}}$)~\cite{NEMO} \\
 & & 1.0$\times10^{32}$($\mathrm{{^{16}O}}$)~\cite{KAMIOKANDE} \\
 $\mathrm{{^{12}C}\rightarrow{^{11}\widetilde{B}}}+ \it{p}$ & 1.0$\times10^{32}$ & 1.9$\times10^{25}$(${^{23}\mathrm{Na}},\!{^{127}\mathrm{\!I}}$)\cite{DAMA09} \\
 $\mathrm{{^{12}C}\rightarrow{^{11}\widetilde{C}}}+ \it{n}$ & 2.0$\times10^{31}$ & 2.1$\times10^{22}$($^{nat}\mathrm{Pb}$)~\cite{Kishimoto} \\
 $\!\!\mathrm{{^{12}\!C}\!\rightarrow\!{^{12}\!\widetilde{N}}}\!+\!\it{e^-}\!\!+\!\overline{\nu}_e$ & 6.4$\times10^{30}$ & 9.5$\times10^{27}$($^{12}\mathrm{C}$)~\cite{Kek90,LSD}\\
 $\!\!\mathrm{{^{12}\!C}\!\rightarrow\!{^{12}\!\widetilde{B}}}\!+\!\it{e^+}\!\!+\!\nu_e$ & 6.6$\times10^{30}$ & 2.6$\times10^{24}$($^{12}\mathrm{C}$)~\cite{NEMO} \\
 \hline
 \end{tabular}
\end{center}
\end{table}

The present analysis uses data that is approximately 10 times larger than the exposure in our previous paper~\cite{Bor10}, which typically yields a threefold increase in sensitivity. 
In fact, for PEP-forbidden processes with neutron and proton emission this resulted in sensitivity increases of 6 and 100 times, respectively. 
The increase of the sensitivity in the case of proton emission is primarily due to the use of MLP-based pulse-shape discrimination, which significantly reduced the background level in the energy region where a proton signal is expected. 
In the case of neutron emission, two correlated events were selected: the first signal from recoil protons and the second from 2.2 MeV $\gamma$-ray. 
Additionally, the prompt event must be a proton-like event. 
In the case of $\beta^\pm$ transitions the upper limit on the lifetime was increased only by a factor of two, since three events appeared in the empty region above 12.5 MeV.
Additionally, the detection efficiency was updated slightly by new improved MC simulations.

\subsection{Limits on the relative strength of PEP-forbidden transitions}
%%%%%%%%%%%%%%%%%%%%%%%%%%%%%%%%%%%%%%%%%%%%%%%%%%%%%%%%%%%%%%%%%%%%%%%%%%%%%%%%%%%%%%%%%%%%%%%%%%%%%%%%%%%%%%%%%%%%%%%%%
Non-Paulian transitions with emission of $\gamma$-quanta, $n$ and $p$ nucleons, and $\beta^\pm$ particles can be induced by electromagnetic, strong, and weak interactions, respectively. 
The obtained above upper limits on lifetime for PEP-forbidden processes ($\tau_{\gamma,p,n,\beta^\pm}$) can be compared with the characteristic lifetime ($\tau$) of corresponding normal processes. 
A measure of the relative strength of non-Paulian transitions to normal transitions is the ratio: $\delta^2$= $\widetilde{\lambda}/\lambda$, where
${\lambda} = 1/\tau$ is the probability per unit time for normal transitions and $\widetilde{\lambda}$ for PEP-forbidden transitions.

The ratio $\delta^2$ %=$(g_{PV}/g_{NT})^2$is a measure of the violation of the PEP and
represents the mixing probability of non-fermion statistics that would allow transitions to the occupied states. 
In particular, in the quon model of PEP-violation~\cite{Ign87,Gre90,Moh90} the parameter $\delta^2 = \beta^2/2$ corresponds to the probability of admixed symmetric component of the particle. 
This approach allows comparison the experimental lifetime limits obtained for different nuclei and atoms.

The single-particle estimate for the partial radiative width of nuclear 
%gamma decays The decay width of the nuclear electric dipole
16.4 MeV E1 $\gamma$-transition from $P$ to $S$ shell, given by the Weisskopf formula ($\mathrm{\Gamma_{\gamma}}\sim E_\gamma^{2L+1} A^{2L/3}$, where A is the mass number and L the multipolarity), is $\mathrm{\Gamma_{\gamma}}\approx$ 1.5~keV. The corresponding rate for a normal E1 transition is then $\lambda = \mathrm{\Gamma_{\gamma}}/ \hbar$=
2.31$\times$10$^{18}$ s$^{-1}$. 
Given the number of nucleons that can undergo a $\gamma$-transition ($N_n=8$) and the obtained upper limit on $\tau_{\gamma}$ (\ref{GLimit}), the ratio
$\delta^2_{\gamma}=\widetilde{\lambda}(^{12}\mathrm{C})/\lambda(^{12}\mathrm{C}) $ is less than 1.04$\times 10^{-57}$ (90~\% C.L.). 
The result obtained with the Kamiokande detector for $^{16}$O nuclei is slightly weaker: $\delta^2_{\gamma}\leq 2.3\times10^{-57}$~\cite{KAMIOKANDE}.

%Although the dipole electric E1-transition is the fastest among $\gamma$-transitions, the width of hadron emissions is 3-4 orders larger than that of $\gamma$-transitions. 
The width of hadronic $(p,n)$ emissions is three to four orders of magnitude larger than that of the fastest dipole electric E1 transition.
The widths of single S-hole states in $\mathrm{^{12}C}$, measured in $(p,2p)$ and $(p,pn)$ reactions, are $\mathrm{\Gamma_{\it{n,p}}} \cong$~12~MeV~\cite{PNPI}, corresponding to a rate $\lambda \cong 1.8\times10^{22}~\mathrm{s^{-1}}$. 
%As a result, the detection 
The study of PEP-forbidden transitions with $p$- or $n$-emissions gives a more stringent limit on the relative strength than the study of $\gamma$-emission if one can obtain a similar limit on the lifetime for both processes. 
Using the conservative lower limits on the lifetimes $\tau_p$ and $\tau_n$ from Eq.~(\ref{limp1}) and (\ref{limn}), respectively, we obtain $\delta^2_{p}=\widetilde{\lambda}/\lambda \leq $
1.39$\times 10^{-60}$ and $\delta^2_{n}\leq $ 7.0$\times 10^{-61}$ at 90\% C.L. 
This result is more than four orders of magnitude stronger than that obtained by the DAMA collaboration~\cite{DAMA09}.

The $\beta^{\pm}$ non-Paulian transitions are first-forbidden $P\rightarrow S$ transitions. 
The characteristic $\mathrm{log}\it{(ft_{1/2})}$ value for such first forbidden transitions is 7.5$\pm$1.5. Taking the conservative value $\mathrm{log}\it{(ft_{1/2})}$=9 yields a lifetime of $\tau\approx 480$~s for $\beta^-$-decay with $Q$ = 18.9~MeV (corresponding to a level width $\mathrm{\Gamma}_{\beta^-}\approx$ 1.4$\times10^{-18}$~eV) and $\tau\approx$1050~s for $\beta^+$ decay with $Q$ = 17.8~MeV. 
Consequently, the limits on the relative strength of non-Paulian $\beta^{\pm}$ decays are significantly weaker: $\delta^2_{\beta^-}\leq $ 9.52$\times 10^{-36}$ and
$\delta^2_{\beta^+}\leq $ 2.01$\times 10^{-35}$ (90\%~C.L.). 
The previous limit $\delta^2_{\beta^-}\leq $ 6.5$\times 10^{-34}$ obtained in Ref.~\cite{Kek90} with LSD data~\cite{LSD} is a factor of 30 weaker. 

It should be noted that although the limits for $\beta^{\pm}$ transitions are significantly weaker than those for nucleon emission, there is a fundamental difference between these processes. 
As mentioned above, in $\beta^{\pm}$ decays a new particle ($p$ or $n$) arises in non-Paulian state; thus the Amado-Primakoff arguments may not apply~\cite{Ama80,Kek90}.
However, there are objections that the newly formed proton or neutron does not leave the nucleus and remains inside the localized system~\cite{Ell2012}.
%However, there is a possibility that since the newly created proton or neutron does not leave the localized nuclear system, the mentioned above principle remains valid. \cite{Ell2012}.
The obtained limits $\delta^2_{\beta^{\pm}}$ can be compared with the limit from the VIP experiment: $\beta^2/2 \leq$8.5$\times 10^{-31}$ for the scattering case and $\beta^2/2 \leq$2.47$\times 10^{-43}$ the close encounters case~\cite{Por2024, Por2025}. %$\beta^2/2 \leq$4.5$\times 10^{-28}$~\cite{Bar06}.
%for electron diffusion model with encounters is $\beta^2/2 \leq \mathrm{2.47\times 10^{-43}}$ for 90\% C.L. \cite{Mil2018, Por2024, Por2025}.

%The limit on sensitivity to of VIP experiment depends on the frequency of electron scattering processes, the sensitivity of transition method depends on the value $\lambda$ which is the frequency testing of lower level by nucleon.

The upper limits obtained on the relative strengths of non-Paulian transitions are shown in Table~\ref{RelStr}. 
For transitions with ($n,p$)- and $\beta^{\pm}$-emission the stronger limit is included.

\begin{table}[!htbp]
\begin{center}
\caption{Upper limits on the relative strength, $\delta^2=\widetilde{\lambda}/\lambda$ (at 90\% C.L.), for non-Paulian
transitions in the Borexino.} \label{RelStr}
\begin{tabular}{|l|c|c|c|c|}
 \hline
 $\!\!\mathrm{mode}$ & $\widetilde{\lambda}(^{12}\mathrm{C})$, & $\lambda(^{12}\mathrm{C})$ & $\delta^2\!=\!\widetilde{\lambda}/\lambda$ & Previous \\
 & (s$^{-1}$) & (s$^{-1}$) & &non-Bx limits \\ \hline
 $\!\gamma$ & $\!\!2.4\!\times\!10^{-39}\!$ & $\!\!2.3\!\times\!10^{18}\!$ & $\!\!\!1.0\!\times\!10^{-57}\!$ & $\!\!2.3\!\times\!10^{-57}\!$\cite{KAMIOKANDE} \\
 $\!n,p\!$ & $\!\!1.3\!\times\!10^{-38}\!$ & $\!\!1.8\!\times\!10^{22}\!$ & $\!\!\!7.0\!\times\!10^{-61}\!$ & $\!\!3.5\!\times\!10^{-55}\!$\cite{DAMA09} \\
 $\!e,\nu\!$ & $\!\!1.9\!\times\!10^{-38}\!$ & $\!\!2.0\!\times\!10^{-3}\!$ & $\!\!\!9.6\!\times\!10^{-36}\!$ & $\!\!6.5\!\times\!10^{-34}\!$\cite{Kek90}\\ %LSD
 \hline
 \end{tabular}
\end{center}
\end{table}

\section{Conclusions}
%%%%%%%%%%%%%%%%%%%%%%%%%%%%%%%%%%%%%%%%%%%%%%%%%%%%%%%%%%
This paper presents improved limits on the probability of Pauli Exclusion Principle violation in nuclei.
Using the unique features of the Borexino detector -- extremely low background, large scintillator mass (278 tons), low energy threshold, and a carefully designed muon-veto system -- we have obtained the following new limits on non-Paulian transitions of nucleons from the $1P_{3/2}$ shell to the $1S_{1/2}$ shell in $^{12}$C with the emission of $\gamma, n, p$ and $\beta^{\pm}$ particles:
\begin{center}
\noindent $\tau(^{12}$C$\rightarrow^{12}\widetilde{\mathrm{C}}+\gamma) \geq 1.1\times10^{32}$~y,
\noindent $\tau(^{12}$C$\rightarrow^{11}\widetilde{\mathrm{B}}+ p) \geq 1.0\times10^{31}$~y,
\noindent
 $\tau(^{12}$C$\rightarrow^{11}\widetilde{\mathrm{C}}+ n) \geq 2.0 \times10^{31}$~y,
\noindent $\tau(^{12}$C$\rightarrow^{12}\widetilde{\mathrm{N}}+ e^- + \nu) \geq 6.4\times10^{30}$~y
%\noindent
\end{center}
and \begin{center} \noindent $\tau(^{12}\mathrm{C}\rightarrow^{12}\widetilde{\mathrm{B}}+ e^+ + \nu_e) \geq 6.6\times10^{30}$ y,
\end{center}
all with 90\% C.L.

Compared with the data of Table~\ref{Lim}, these limits for non-Paulian transitions in $^{12}$C with $\gamma$, $p$, $n$, and $\beta^{\pm}$ emission are the most stringent to date. 
These lifetime limits yield new upper limits on the relative strengths of non-Paulian to normal transitions: $\delta^2_{\gamma}\leq 1.0\times 10^{-57}$, $\delta^2_{N}\leq 7.0\times 10^{-61}$, and $\delta^2_{\beta}\leq 9.6\times 10^{-36}$, all at 90\% C.L.

\section{Acknowledgments}
%%%%%%%%%%%%%%%%%%%%%%%%%%%%%%%%%%%%%%%%%%%%%%%%%%%%%%%%%%%%%%
The Borexino program was made possible by funding from Istituto Nazionale di Fisica Nucleare (Italy), National Science Foundation (USA), Russian Science Foundation (Grant No. 24-12-00046) (Russia), Deutsche Forschungsgemeinschaft, Cluster of Excellence PRISMA+, and recruitment initiative of Helmholtz-Gemeinschaft (Germany), and Narodowe Centrum Nauki (Poland). 
We acknowledge the generous hospitality and support of the Laboratori Nazionali del Gran Sasso (Italy).

\end{document}